\documentclass[twocolumn,trackchanges,tighten]{aastex62}

\usepackage{multirow}
\usepackage[normalem]{ulem}	
\usepackage{xspace}
\usepackage{amsmath}
\usepackage{bm}
\newcommand{\ulcnmcurnthreec}{9.2(6) \times 10^{8}}
\newcommand{\ulcnmhdthreec}{9.3(2) \times 10^{8}}

\newcommand{\ptarcade}{{\tt PTArcade}\xspace}

\DeclareMathOperator{\Tr}{Tr}

\defcitealias{NG11a}{NG11a}
\defcitealias{NG12p5data}{NG12p5data}
\defcitealias{NG12p5gwb}{NG12p5gwb}

\graphicspath{{./}{./figures/}}

\submitjournal{ApJL}
\shorttitle{The NANOGrav 15 yr Data Set: Impacts of CNM on GW analyses}
\shortauthors{The NANOGrav Collaboration}

\begin{document}

\title{The NANOGrav $15$ yr Data Set: Impacts of Customized\\ Chromatic Noise Models on Gravitational Wave Analyses}

\correspondingauthor{Jeremy G. Baier}
\email{jeremy.baier@nanograv.org}
\correspondingauthor{Bjorn Larsen}
\email{bjorn.larsen@nanograv.org}
\author[0000-0001-7544-7876]{Nikita Agarwal}
\affiliation{Department of Physics and Astronomy, West Virginia University, P.O. Box 6315, Morgantown, WV 26506, USA}
\author[0000-0001-5134-3925]{Gabriella Agazie}
\affiliation{Center for Gravitation, Cosmology and Astrophysics, Department of Physics and Astronomy, University of Wisconsin-Milwaukee,\\ P.O. Box 413, Milwaukee, WI 53201, USA}
\affiliation{Center for Gravitational Waves and Cosmology, West Virginia University, Chestnut Ridge Research Building, Morgantown, WV 26505, USA}
\author[0009-0008-9486-8463]{Alessandra Amosso}
\affiliation{Institute for Theoretical Physics, University of M\"{u}nster, 48149 M\"{u}nster, Germany}
\author[0000-0002-8935-9882]{Akash Anumarlapudi}
\affiliation{Department of Physics and Astronomy, University of North Carolina, Chapel Hill, NC 27599, USA}
\author[0000-0003-0638-3340]{Anne M. Archibald}
\affiliation{Newcastle University, NE1 7RU, UK}
\author[0009-0008-6187-8753]{Zaven Arzoumanian}
\affiliation{X-Ray Astrophysics Laboratory, NASA Goddard Space Flight Center, Code 662, Greenbelt, MD 20771, USA}
\author[0000-0002-8395-957X]{Anjana Ashok}
\affiliation{Department of Physics, Oregon State University, Corvallis, OR 97331, USA}
\altaffiliation{NANOGrav Physics Frontiers Center Postdoctoral Fellow}
\author[0000-0002-4972-1525]{Jeremy G. Baier}
\affiliation{Department of Physics, Oregon State University, Corvallis, OR 97331, USA}
\author[0000-0003-2745-753X]{Paul T. Baker}
\affiliation{Department of Physics and Astronomy, Widener University, One University Place, Chester, PA 19013, USA}
\author[0000-0003-0909-5563]{Bence B\'{e}csy}
\affiliation{Institute for Gravitational Wave Astronomy and School of Physics and Astronomy, University of Birmingham, Edgbaston, Birmingham B15 2TT, UK}
% ...



\begin{abstract}
\label{sec:abstract}

We report updated nHz gravitational wave (GW) significance, characterization, and interpretations using the customized chromatic-noise models (CNMs) developed in \citet{NG15_CNM} for the NANOGrav 15-year data set. We find increased evidence for the Hellings-Downs (HD) correlation signature of the stochastic gravitational wave background (GWB), with a Bayes factor of $1571\pm14$ for HD-correlations over a common uncorrelated red-noise process using a power-law model with $14$ Fourier modes. We find this $\sim8\times$ increase in Bayes factor from \citet{ng15gwb} is a result of improved noise mitigation. Assuming an analytic null distribution for the frequentist interpulsar correlation statistic, this corresponds to a slightly more significant measurement from $3.16\sigma$ to $3.32\sigma$ against the no-correlation scenario. Spectral inference with CNMs brings the power-law GWB amplitude down to $A_{\rm GWB}=2.1^{+0.6}_{-0.5}\times10^{-15}$ at fixed $\gamma_{\rm GWB}=13/3$. In a varied-$\gamma$ analysis, the spectral index increases to $\gamma_{\rm GWB}=3.5^{+0.7}_{-0.6}$. We report updates on an all-sky continuous gravitational wave (CW) search as well as select targeted searches and calculate a $3.2\times$ larger detection volume for the NANOGrav detector. With CNMs, we find reduced evidence for a non-Einsteinian, scalar-transverse mode of gravity. Finally, we reinterpret the GWB first with the assumption of an astrophysical background sourced by SMBHBs and then assuming the more exotic origins of cosmic inflation, a first-order cosmological phase transition, and stable cosmic strings. Under both the SMBHB hypothesis and the cosmological hypotheses, we see only marginal shifts in model parameter posteriors which are consistent with the slightly quieter and steeper power-law GWB spectrum.

\end{abstract}

\section{Introduction} 
\label{sec:intro}
Pulsar timing arrays (PTAs) rely on the rotational stability of millisecond pulsars to form a galactic gravitational-wave (GW) detector \citep{Detweiler:1979wn, hd83}. By correlating pulse times of arrival (TOAs) across the array, PTAs are sensitive to GWs in the nanohertz frequency band. A stochastic gravitational wave background (GWB), expected from the cosmic population of supermassive black hole binaries (SMBHBs), should dominate this band \citep{R&R1995, Jaffe+2003}. The community of international PTA collaborations have recently reported varying evidence for a GWB emerging in their datasets \citep{ng15gwb,eptadr2_3:gwb,pptadr3:gwb, cptadr1_1:gwb, MPTADR2_noise_and_gwb, ipta3p+2024}.

Since PTA datasets are built inclusively from previous data releases, i.e. datasets continue to get longer, with more TOAs, the signal significance is expected to grow slowly over time \citep{siemens+2013, hazboun:2020slice}. At this stage of significance for nHz GWs, PTAs remain noise-model dependent, which presents a key challenge in inference and interpretation of any emerging signal. Sources of noise in PTAs include intrinsic spin noise, variations in the interstellar medium (ISM) that cause dispersion-measure (DM) and scattering fluctuations, time-variable solar-wind contributions to DM, pulse phase and profile jitter, band-/system-dependent noise, and array-wide contributions such as clock errors and Solar System ephemeris uncertainties \citep{tiburzi+2016,ng15detchar}. Mis-modeling any of these can bias the inference of a common-spectrum process, potentially inflating or suppressing Hellings and Downs correlations as well as polluting the inferred GW spectrum \citep{DiMarco2024}. In particular, the solar wind has been shown to be an important common chromatic signal that needs to be carefully modeled, in concert with other dispersion variations, to effectively remove long-time scale and short-time scale variations in dispersion \citep{hazboun+2022sw,susarla+24sw,ng12p5_CNM,Iraci+2025,NG15_CNM}. 

Gaussian process modeling \citep{rw06} for pulsar timing data has been in use from early on in the field \citep{vanhaasteren+2014_gp, lentati+:2016}, where Fourier-domain kernels are especially useful for modeling the various power-law red noise physics that occurs in PTA datasets, e.g. DM variations and the GWB.
\citet{ng12p5_CNM} introduced a number of time-domain kernels for chromatic noise processes that were shown  to also successfully mitigate noise. This analysis used custom models for each pulsar analyzed, a tactic that has been adopted fairly universally by PTAs \citep{lentati+:2016, goncharov+2021, Chalumeau+2022, Larsen2024, MPTA_DR2_noise, inpta_dr2_noise}. Independent, full-PTA studies with separate datasets \citep{hazboun:2020slice, pptadr3:noise, MPTADR2_noise_and_gwb, Goncharov+2025, ng12p5_CNM} have consistently found that improved treatment of pulsar noise shifts the common-spectrum amplitude and slope (typically to a quieter and steeper power-law spectrum) relative to simpler models. 

The spectral characterization of the GWB is an invaluable product of PTA GW searches that is used by the SMBHB community and the new physics community as the input for various types of source classification and characterization \citep{ng15smbbh,ng15newphysics}. \citet{sato-polito2024} have argued that the GWB amplitude inferred by PTAs may be too large to arise solely from SMBHBs, unless SMBH masses have been under-estimated. Our present analysis directly explores this possibility by examining how robust the inferred GWB amplitude remains under alternative noise assumptions. This effect of noise models is not only central to stochastic background detection but also impacts other PTA science goals. For continuous-wave (CW) searches, the robustness of noise models is essential to avoid misidentifying individual SMBHBs \citep{Falxa+2023, ng15singlesource, ng15-targeted}. Likewise, excursions in the recovered GWB strain spectrum, either from deficits or excesses of power, may reflect noise artifacts or genuine over/underdensities of SMBHBs in the Universe. \citet{ng15-discreteness} demonstrated this in the strain spectrum analysis of the NANOGrav 15 yr dataset, where distinguishing noise features from astrophysical structure becomes critical. 
At the time of writing, \citet{ng15gwb} has over $1000$ citations, many of those studies use the posteriors or point estimates from \citet{ng15gwb} as the starting point for theoretical motivation and/or model inputs for both astrophysical population modeling and cosmological ``new-physics'' studies. Consequently, changes in spectral characterization have varied and far reaching impacts.

In this work, we present how customized, chromatic-noise model selection over all $67$ pulsars in the NANOGrav 15 year data set (NG15) impacts GW results. To this end, we reanalyze key GW results presented in \citet{ng15gwb, ng15singlesource, ng15detchar, ng15smbbh, ng15_altpol, ng15_ppc, ng15-targeted, ng15newphysics} with the CNM developed in \citet{NG15_CNM}. We show that the significance for a GWB is stronger using CNM and spectral characterization is in less tension with the galaxy population models in \citet{sato-polito2024}.  We further show how custom noise models mitigate spurious signals, particularly those in searches for alternative GW polarizations and CWs. The overall astrophysical and cosmological interpretations remain largely unchanged since they are currently prior dominated.

This paper is organized as follows: In \S\ref{sec:methods} we recap the methods used in the NANOGrav GW search papers, and introduce a new computational technique in \S\ref{sec:cached-chromatic-likelihood} for efficiently sampling a likelihood that includes the many chromatic noise models we use herein. In \S\ref{sec:results} we present the various GW searches we have reproduced using custom chromatic noise models and compare them to results using the standard noise model. In \S\ref{sec:astro_cosmo_interp} we present the cosmological and astrophysical interpretation results from \citet{ng15smbbh} and \citet{ng15newphysics}, again using CNM. In \S\ref{sec:conclusions} we make a number of concluding remarks about the use of chromatic noise models in the future. 

\section{Analysis Methods}\label{sec:methods}
\label{sec:methods}

Here we describe our analysis methods, which are largely the same as past analyses, with the exception of a new, faster likelihood implementation for our specific analysis in \S\ref{sec:cached-chromatic-likelihood}. \S\ref{sec:noise-models} gives a summary of the pulsar noise models and how they are applied during GW analyses. \S\ref{sec:likelihood} recapitulates the PTA likelihood, with \S\ref{sec:cached-chromatic-likelihood} introducing our implementation difference for fixed GP noise parameters. \S\ref{sec:meth_bf} describes our Bayes Factor calculation methods. \S\ref{sec:gwbmeth} summarizes the GWB search models, with details on the PTA optimal statistic in \S\ref{sec:OS_methods}. Finally, \S\ref{sec:cwmeth} details the CW search methods.

\subsection{Noise Models}
\label{sec:noise-models}
The chromatic-noise models used in this work are presented and thoroughly studied in \citet{NG15_CNM}. The major difference between the noise models presented in \citet{ng15detchar} and \citet{NG15_CNM} is that the ``DMX'' model for changes in dispersion measure is replaced by a Gaussian process model (DMGP) and other chromatic signals are tested for in a customization process. Accordingly, we refer to these models as the custom noise models (CNM); however, the customization process only applies to the chromatic (i.e. radio frequency dependent) portion of the model and leaves the achromatic portion identical to those in \citep{ng15detchar}. The noise models are customized using the criterion of Bayesian evidence and select for which \emph{chromatic} signals, which basis and covariance, and which basis resolution are appropriate for each pulsar's line of sight through the interstellar and interplanetary media. A global, deterministic solar-wind model is fit for and applied to all pulsars but select pulsars have a stochastic solar-wind model on top of that. Additional models for yearly and transient chromatic variations are tested for and applied where deemed significant. See \citet{NG15_CNM} for more details on the CNMs, model selection process, and model validation.

Previous analyses have found chromatic noise parameters can be covariant with achromatic noise parameters, and by extension with common GW signals \citep{lentati+:2016, eptadr2_2:noise, MPTADR2_noise_and_gwb, Larsen+2025}, particularly if radio frequency band coverage is poor \citep{Ferranti+2025_bw}. Nonetheless, analysis of the NG15 single-pulsar noise posteriors in \citet{NG15_CNM} showed that the DM, free chromatic, and SW parameters are not covariant with achromatic red noise parameters for all but two pulsars, indicating that most NG15 pulsars have been observed with sufficient radio frequency coverage to overcome this confusion between chromatic and achromatic noise sources. For this reason, we follow the approach from \citet{ng12p5_CNM} to carry out GW analyses with the majority of CNM parameters set to their \emph{maximum a posteriori} (MAP) values from the single pulsar noise analyses, which allows many of these analyses herein to become computationally tractable. To test the validity of this approach, we also perform select GW analyses (where indicated in \S\ref{sec:results}) with the chromatic parameters allowed to vary (this requires parallel tempering during the MCMC analysis to achieve convergence; \citealt{Johnson+2024}). We find that fixing the CNM parameters may result in some loss of GW sensitivity, but the change is not large enough to impact the conclusions of this work. At the very least, the impact should be no greater than the common practice of fixing the single-pulsar achromatic white noise parameters \citep{ng15detchar}, noting however that the interplay between white noise and CW signals is still an ongoing challenge within the PTA community \citep{IPTA_MDC2}.

At the outset of our fixed-point analyses, we decided to fix all CNM parameters \emph{except for} PSR J1713+0747's exponential event parameters since previous work suggested these two events may have an outsized impact on GW analyses with NG15 \citep{hazboun:2020slice, Larsen2024}. These events show associated profile variability \citep{Lin+2021, goncharov+2021} and are described in detail with the interpretation of an ISM-origin by \citet{lam+2018}, although a more dramatic and recent chromatic event (beyond the timespan of NG15) suggests some undetermined phenomena within the pulsar magnetosphere as a likelier origin for the events \citep{jennings+2022, Mandow+2025}. We ultimately find that even though the inclusion of the two exponential event models in the CNMs is indeed impactful, the exponential event posteriors show no significant changes between the single-pulsar and full-PTA analyses.

\subsection{Likelihood}
\label{sec:likelihood}

Here we recap the traditional construction of the PTA likelihood for GW searches used in \citet{ng15gwb, ng15detchar}, and note the slight differences in implementation required to run analyses with our per-pulsar chromatic-noise models. Likelihoods and priors throughout this work are constructed using \texttt{enterprise} \citep{enterprise}. We also present a new modification to the likelihood which is more efficient when assuming fixed noise hyperparameters.

\subsubsection{Traditional Likelihood}

The log likelihood is defined
\begin{align}
    \log\mathcal{L} = -\frac{1}{2}\bm{r}^T\bm{C}^{-1}\bm{r} - \frac{1}{2}\log\det(2\pi\bm{C}).
\end{align}
The residual vector is defined
\begin{equation}
    \bm{r} = \bm{\delta t} - \bm{s}(\bm{\theta}_{\rm det}).
\end{equation}
This is just our original timing residuals $\bm{\delta t}$ with a deterministic signal vector $\bm{s}$ depending on parameters $\bm{\theta}_{\rm det}$ subtracted. In this work, only continuous GWs and per-pulsar chromatic events are treated as deterministic. The covariance matrix $\bm{C}$ contains all remaining stochastic signals and noise, including the GWB. To compute $\bm{C}^{-1}$ most efficiently, we use the two-step marginalization process from \citet{Johnson+2024}, in which the covariance matrix is first split up as
\begin{align}
    \bm{C} &= \bm{D} + \bm{\Sigma},
\end{align}
with $\bm{D}$ and $\bm{\Sigma}$ defined 
\begin{align}
    \bm{D} &= \bm{N} + \bm{M}\bm{E}\bm{M}^T, \\
    \bm{\Sigma} &= \bm{F}\bm{\Phi}\bm{F}^T.
\end{align}

In the $\bm{D}$ matrix, $\bm{N}$ is a block-diagonal, time-uncorrelated noise matrix including contributions of TOA measurement uncertainties, EFAC, EQUAD, and ECORR. $\bm{M}$ is the timing model design matrix (computed after expanding the timing model to linear order in the perturbations to the best-fit timing model parameters $\bm{\beta}_0$, $\bm{\epsilon} \equiv \bm{\beta} - \bm{\beta}_0$), and $\bm{E} = \langle\bm{\epsilon}\bm{\epsilon}^T\rangle$ is the prior uncertainty on the linear timing model perturbations. As the timing model is fit using an unconstrained least-squares regression and is assumed to be data-dominated rather than prior-dominated, by convention we set $\bm{E} \to 10^{40}\bm{I}$ to approximate infinite prior variance, such that $\bm{E}^{-1} \approx \bm{0}$. Via the Woodbury matrix identity, the inverse is evaluated as
\begin{align}
    \bm{D}^{-1} &= \bm{N}^{-1} + \bm{N}^{-1}\bm{M}(\bm{M}^T\bm{N}^{-1}\bm{M})^{-1}\bm{M}^T\bm{N}^{-1},
\end{align}
which we then cache in its entirety as by convention we assume all per-pulsar white noise parameters (EFAC, EQUAD, and ECORR) are held fixed to their \emph{maximum a posteriori} (MAP) values from single pulsar analyses. In \citet{ng15gwb, ng15detchar}, all DM variations are featured in $\bm{M}$ as a series of piecewise constant functions for DM represented by the DMX parameters in \texttt{PINT} \citep{PINT}. In our analysis, all DMX columns are removed from $\bm{M}$, and 3 columns are added to the design matrix corresponding to the DM, DM1, DM2 parameters, filtering a quadratic trend in DM.

Meanwhile, $\bm{\Sigma}$ contains stochastic red noise contributions, including contributions from either a CURN or a GWB, where $\bm{F}$ is the total design matrix (which maps the timing residuals onto a low-dimensional latent space) and $\bm{\Phi}$ is the covariance matrix in the latent space. The inverse is again computed via the Woodbury matrix identity as
\begin{align}
    \bm{C}^{-1} &= \bm{D}^{-1} + \bm{D}^{-1}\bm{F}\bm{\Theta}\bm{F}^T\bm{D}^{-1},
\end{align}
where
\begin{align}
    \bm{\Theta} = (\bm{\Phi}^{-1} + \bm{F}^T\bm{D}^{-1}\bm{F})^{-1}.
    \label{eq:Theta_mat}
\end{align}
To further improve computational performance, common and intrinsic red noise components typically share a truncated sine-cosine Fourier basis $\bm{F}^{\rm RN}$, such that the corresponding block of the covariance matrix $\bm{\Phi}^{\rm RN}$ is $N_{\rm psr}\times 2N_f$. Using a model which is purely auto-correlated (i.e., the CURN), $\bm{\Phi}^{\rm RN}$ is diagonal and trivial to invert, whereas a model with interpulsar correlations included (i.e. the GWB), will feature off-diagonal elements in $\bm{\Phi}^{\rm RN}$.

Following standard convention in \texttt{enterprise}, our analysis includes the contributions from DM, solar wind, and free chromatic variations as an expansion to the latent space within $\bm{\Sigma}$ as
\begin{align}
    \bm{F} &= [\bm{F}^{\rm RN}, \bm{F}^{\rm DM}, \bm{F}^{\rm FC},\bm{F}^{\rm SW}], \\
    \bm{\Phi} &= [\bm{\Phi}^{\rm RN}, \bm{\phi}^{\rm DM}, \bm{\phi}^{\rm FC},\bm{\phi}^{\rm SW}],
\end{align}
where lowercase $\bm{\phi}$ is distinguished to indicate noise sub-blocks whose components are uncorrelated between pulsars, whereas $\bm{\Phi}^{\rm RN}$ is the GW+RN sub-block whose components will be correlated between pulsars when modeling a GWB. Not all pulsars contain solar wind and free chromatic contributions; each pulsar actually uses it own customized set of chromatic design matrices, as determined in \citet{NG15_CNM}. Furthermore, some pulsars favor the coarse-grained time-domain design matrices for chromatic noise introduced in \citet{ng12p5_CNM}, which do not necessarily yield diagonal $\bm{\phi}$ matrices. Using this version of the likelihood, coupling these non-diagonal $\bm{\phi}$ blocks with a non-diagonal $\bm{\Phi}^{\rm RN}$ block unfortunately breaks the ability to compute $\bm{\Phi}^{-1}$ directly using the sparse matrix methods in \texttt{enterprise}, instead requiring dense matrix operations \citep{Johnson+2024}. This fact, alongside the significantly increased number of elements in $\bm{\Sigma}$, means the time to evaluate $\bm{\Phi}^{-1}$ and $\bm{\Theta}$ is significantly longer than is usual when using the standard noise models.

\vspace{\baselineskip}

\subsubsection{Cached-Chromatic Likelihood}
\label{sec:cached-chromatic-likelihood}
We are able to circumvent the above issues when assuming \emph{fixed} chromatic noise parameters (as discussed above in \S\ref{sec:noise-models}) by extending the two-step marginalization process from \citet{Johnson+2024} to an $N+2$-step marginalization, defining
\begin{align}
    \bm{D} &= \bm{N} + \bm{M}\bm{E}\bm{M}^T + \sum_k^N \bm{\Sigma}_k, \\
    \bm{\Sigma}_k &= \bm{F}_k\bm{\phi}_k\bm{F}_k^T,
\end{align}
where $N$ is the number of additional fixed-hyperparameter processes and $\bm{F}_i$, $\bm{\phi}_i$ represent the GP design matrix and latent-space covariance matrix of the $k$-th process. Suppose we define a function for the matrix inverse from the Woodbury matrix identity as
\begin{align}
    \mathcal{I}(\bm{N},\bm{\Sigma}_k) &\equiv \bm{N}^{-1} + \bm{N}^{-1}\bm{F}_k\bm{\Theta}_k\bm{F}_k^T\bm{N}^{-1}, \\
    \bm{\Theta}_k &= (\bm{\phi}_k^{-1} + \bm{F}_k^T\bm{N}^{-1}\bm{F}_k)^{-1}.
\end{align}
The inverse of the expanded $\bm{D}$ matrix may be obtained recursively, where the inverse is recomputed using the previous output from $\mathcal{I}$. In function composition notation, this process is defined compactly as
\begin{align}
    \bm{D}^{-1} = \mathcal{I}(\cdot,\bm{\Sigma}_k) \circ \cdots \circ \mathcal{I}(\cdot,\bm{\Sigma}_1)\circ \mathcal{I}(\bm{N},\bm{M}\bm{E}\bm{M}^T).
\end{align}
This allows all DM, free chromatic, and solar wind components to be added one at a time to the timing and white noise processes in the total $\bm{D}$ matrix, leaving only GW and intrinsic red noise contributions in the remaining $\bm{F}$ and $\bm{\Phi}$ matrices as originally the case in \citet{ng15gwb, ng15detchar}. So long as all chromatic GP hyperparameters are held fixed, $\bm{D}^{-1}$ can be cached in its entirety, allowing more efficient likelihood evaluations. $\bm{\Phi}^{-1}$ also becomes easier to compute as the non-GW elements of $\bm{\Phi}$ are fully-diagonal. Using our CNMs with NG15, we find the new likelihood evaluation time (as compared against the traditional likelihood with chromatic terms in the $\bm{\Phi}$ matrix) decreases by a factor of $\sim 2$ when modeling a CURN, and the evaluation time decreases by a factor of $\sim 4$ when including HD correlations. Note in our implementation that the deterministic signal vector $\bm{s}(\bm{\theta}_{\rm det})$ which includes transient exponential events for select pulsars still remains a bottleneck for evaluating the likelihood.

\vspace{\baselineskip}

\subsection{Bayes Factor Estimation}
\label{sec:meth_bf}
In this work, we calculate Bayes factors using three techniques which are well established in GW data analysis. We use product space sampling \citep{Hee+2016, Johnson+2024} where a model switch parameter is inferred and the ratio of MCMC samples in one model compared to another estimates the odds ratio between those models. We also use likelihood reweighting \citep{Hourihane+2023}, which is a form of importance sampling, whereby posterior samples from one model are passed to the likelihood of another model and the average log-likelihood ratio of the models estimates the Bayes factor between the models. Lastly, we use the Savage-Dickey density ratio \citep{Dickey1971} for cases of nested models (e.g. CW searches) wherein a Bayes factor for an outer model over a nested inner model is estimated by computing the ratio of the prior to the posterior density in a region of parameter space where the outer model is effectively equivalent to the inner model.

\subsection{GWB analyses}
\label{sec:gwbmeth}

The GWB models and analysis methods are largely unchanged from previous works \citep{ng15gwb, ng15detchar, ng15_ppc, ng15_altpol, Johnson+2024}. Throughout this section we use indices $a,b$ to denote distinct pulsars, and indices $i,j$ to label distinct TOAs.

GWB signals are treated as common red noise processes in the full PTA. Mathematically, the total red noise covariance matrix is split into $\bm{\Phi}^{\rm RN} = \bm{\Phi}^{\rm CRN} + \bm{\phi}^{\rm IRN}$, where $\bm{\Phi}^{\rm CRN}$ contains the GWB signals and $\bm{\phi}^{\rm IRN}$ is block-diagonal between pulsars and includes just the pulsar-intrinsic contributions. The most common parameterization for $\bm{\Phi}^{\rm CRN}$ is the power law (PL) spectral model,
\begin{align}
    \label{eq:powerlaw_crn}
    \Phi^{\rm CRN,PL}_{ij,ab} &= \frac{A_{\rm CRN}^2}{12\pi^2f_{\rm yr}^3T}\left(\frac{f_i}{f_{\rm yr}}\right)^{-\gamma_{\rm CRN}}\delta_{ij}\Gamma_{ab},
\end{align}
where $A_{\rm CRN}$ and $\gamma_{\rm CRN}$ are hyperparameters which are sampled during the full analysis. The model is motivated as a value of $\gamma = 13/3$ is expected for a population of SMBHBs under the analytical limit \citep{Phinney2001}. Other spectral models we use here include a ``free spectrum'' which is more agnostic to the source of the GWB,
\begin{align}
    \label{eq:free_spectral_prior}
    \Phi^{\rm CRN,FS}_{ij,ab} &= \rho^2(f_i)\delta_{ij}\Gamma_{ab},
\end{align}
where $\rho^2(f_i)$ is a separate hyperparameter at each frequency, or a broken power law (BPL) spectrum,
\begin{align}
    \label{eq:bpl}
    \Phi^{\rm CRN,BPL}_{ij,ab} &= \Phi^{\rm CRN,PL}_{ij,ab}\left[1+\left(\frac{f_i}{f_{b}}\right)^{1/\ell}\right]^{\ell\gamma_{\mathrm{CRN}}},
\end{align}
which allows the spectral shape to flatten out past some break frequency $f_b$, with smoothing parameter $\ell$. See Table~\ref{tab:priors} for priors on these various spectral hyperparameters.

The number of Fourier elements in the model is typically truncated to some low number since the GWB is a low-frequency signal; for the NG15 analysis, $N_f = 14$ (modeling up to $\sim28$ nHz) was selected based on a broken power law analysis \citep{ng15gwb}, but it is typical to test the impact of different values for the truncation frequency. To reduce the risk of mismodeling intrinsic pulsar contributions as common signals, each pulsar's intrinsic red noise sub-block $\bm{\phi}^{\rm RN}$ uses the same spectral model with independent noise parameters $A_a$ and $\gamma_a$ for each pulsar, with intrinsic red noise always modeled out to $N_f = 30$ (or $\sim60$ nHz).

$\Gamma_{ab}$ is the overlap reduction function (ORF) encoding the cross-correlations between pulsar $a$ and $b$. For an isotropic GWB, the ORF is given by the HD curve,
\begin{align}
    \Gamma^{\rm HD}_{ab} &= \frac{3}{2}x_{ab}\ln(x_{ab}) - \frac{1}{4}x_{ab} + \frac{1}{2} + \frac{1}{2}\delta_{ab}, \\
    x_{ab} &= \frac{1 - \cos(\zeta_{ab})}{2},
\end{align}
where $\zeta_{ab}$ is the angular sky separation between pulsars $a$ and $b$. The conservative null hypothesis for the GWB is a common uncorrelated red noise (CURN) process, which is given by the exact same spectral model, Eq.~\eqref{eq:powerlaw_crn}, but with $\Gamma_{ab} \to \delta_{ab}$. This uncorrelated ORF is the likeliest match to the common signal if intrinsic noise modeling errors spuriously and independently arise in multiple pulsars \citep{Zic+2022}. Various modifications to the HD ORFs may also arise either from an anisotropic GWB \citep{mingarelli+2013, taylor+2013_anisotropy}, cosmic variance in the SMBHB population \citep{allen2022, AllenRomano2023}, an individual SMBHB \citep{CornishSesana2013, Schult+2025, Mingarelli+2026_fingerprints}, or alternative (non-Einsteinian) polarization modes of gravity \citep{ChamberlinSiemens2012, Zheng+2026}. We test one such alternative polarization hypothesis from \citet{ng15_altpol} in \S\ref{sec:ST_modes}. Non-GW sources of interpulsar cross-correlations include errors in observatory time standards (monopolar; $\Gamma_{ab} = 1$), solar system ephemerides (dipolar; $\Gamma_{ab} = \cos\zeta_{ab}$), and dispersive errors induced by solar wind (quasi-dipolar, but depends on the pulsar and TOA configuration; \citealt{tiburzi+2016}). See \citet{ng15gwb} for an extensive investigation of monopolar and dipolar correlations using the NG15 dataset.

To sample in the GWB and intrinsic red noise models, we use \texttt{PTMCMCSampler} \citep{ptmcmcsampler}, which has been verified as a reliable tool for Bayesian inference with full PTA datasets \citep{Johnson+2024}. \texttt{PTMCMCSampler} natively runs Markov Chain Monte Carlo (MCMC) using an Adaptive Metropolis algorithm, but allows a slew of other jump proposals, including a parallel tempering algorithm which we use as a cross-check. In particular, we find parallel tempering is required for our analyses which allow the chromatic parameters of our CNMs to vary. When modeling the GWB with HD correlations, we make us of the likelihood reweighting algorithm from \citet{Hourihane+2023} where applicable.

\subsection{Optimal Statistic}
\label{sec:OS_methods}

To complement the likelihood-based analyses, we also use the PTA ``optimal statistic'' (OS; \citealt{anholm+2009os, chamberlin+2015os}), which provides an analytic estimator of the cross-correlated GWB amplitude, 
\begin{align}
    \hat{A}^2 = \frac{\sum_{a>b}\bm{\delta t}_a^T\bm{P}_a^{-1}\bm{\tilde{S}}_{ab}\bm{P}_b^{-1}\bm{\delta t}_b}{\sum_{a>b}\Tr\left[\bm{P}_a^{-1}\bm{\tilde{S}}_{ab}\bm{P}_a^{-1}\bm{\tilde{S}}_{ba}\right]},
\end{align}
and signal-to-noise ratio,
\begin{align}
    {\rm{S/N}} = \frac{\hat{A}^2}{\sigma_0} = \frac{\sum_{a>b}\bm{\delta t}_a^T\bm{P}_a^{-1}\bm{\tilde{S}}_{ab}\bm{P}_b^{-1}\bm{\delta t}_b}{\left(\sum_{a>b}\Tr\left[\bm{P}_a^{-1}\bm{\tilde{S}}_{ab}\bm{P}_a^{-1}\bm{\tilde{S}}_{ba}\right]\right)^{1/2}},
    \label{eq:OS_SNR}
\end{align}
where $a$, $b$ iterate over pulsars, and the time-domain data covariance is separated into the total pulsar auto-covariance $\bm{P}$ and cross-covariance $\bm{S} = \hat{A}^2\bm{\tilde{S}}$. The OS framework has been extended in numerous ways (e.g., \citealt{Vigeland+2018, Sardesai+2023, Gersbach+2025}); particularly, the OS provides a straightforward way to compute a $p$-value (or equivalently, a Gaussian $\sigma$ significance level) for GWB detection with PTAs because as a quadratic form, its null distribution is well-defined analytically as a generalized $\chi^2$ (GX2) distribution \citep{hazboun+2023gx2, vanHaasteren+2025_pvals}.

To assess the impact of custom chromatic noise using the OS, we first use the noise-marginalized OS (NMOS; \citealt{Vigeland+2018}), to reproduce the GWB S/N distribution from \citep{ng15gwb}, numerically marginalized over the pulsar noise posteriors. 
We then repeat the analysis from \citet{ng15_ppc} to obtain Bayesian $p$-values \citep{Vallisneri+2023}, which we compute rapidly using the rank-reduced framework from \citet{vanHaasteren+2025_pvals}. Results of this analysis may be found in \S\ref{sec:OS_results}. Note we do not account for the effects of covariance between pulsar pairs in this analysis, as this covariance assumes the presence of a GWB, which is not proper for a detection statistic \citep{AllenRomano2023, Gersbach+2025}. Additionally, the original analyses we compare with did not include this pair covariance.

\subsection{Continuous Wave Searches}
\label{sec:cwmeth}
Since a GWB sourced by SMBHBs may eventually have resolvable individual sources \citep{rosado+2015}, we perform searches for continuous GWs (CW) across the entire sky and from specific candidate SMBHB systems. In the latter, ``targeted'' CW searches, we fold-in constraints available from electromagnetic surveys, as this can enhance the GW search sensitivity and decreases the computational expense \citep{NANOGrav:2020lwu}.

A CW from direction $\hat{\Omega}$, induces a signal 
\begin{equation}
s(t, \hat{\Omega})=F^{+}(\hat{\Omega}) \Delta s_{+}(t)+F^{\times}(\hat{\Omega}) \Delta s_{\times}(t)
\end{equation}
in a pulsar's timing residuals.
Here $F^{+, \times}(\hat{\Omega})$ are the pulsar's antenna pattern functions and $\Delta s_{+, \times}$, the difference between the ``Earth term" and the ``pulsar term" of the signal. Assuming circular binary orbits,
\begin{equation}
\begin{aligned} 
s_{+}(t)= \frac{\mathcal{M}_c^{5 / 3}}{d_{L} \omega(t)^{1 / 3}}\left[-\sin 2 \Phi(t)\bigl[(1+\cos ^{2} \iota\right) \cos 2\psi \\  - 2 \cos 2\Phi(t) \cos \iota \sin 2 \psi\bigr],
\end{aligned} 
\end{equation}
\begin{equation}
\begin{aligned} 
s_{\times}(t)= \frac{\mathcal{M}_c^{5 / 3}}{d_{L} \omega(t)^{1 / 3}}\bigl[-\sin 2 \Phi(t)\left(1+\cos ^{2} \iota\right) \sin 2\psi \\ + 2 \cos 2\Phi(t) \cos \iota \cos 2 \psi\bigr], 
\end{aligned}
\end{equation}
where $d_L$ is the luminosity distance to the source, $\iota$ its inclination angle,  $\omega(t)$ and $\Phi(t)$ are the time-dependent angular orbital frequency and phase, respectively, and $\mathcal{M}_c$ is the chirp mass. The orbital frequency is 
\begin{equation}
\omega(t) = \omega_0 \left(1 - \frac{256}{5} \mathcal{M}_c^{\frac{5}{3}} \omega_0^{\frac{8}{3}} (t-t_0)\right)^{-\frac{3}{8}},
\end{equation}
given orbital frequency $\omega_0$ at a reference epoch $t_0$. 

For the case of targeted CW searches, we assume that the GW frequency is twice the orbital frequency which is estimated in EM observations. We then use a Dirac delta prior on the GW frequency as well as the sky location parameters and luminosity distance \citep{arz:2020, ng15-targeted}.

We note that in all CW searches, we include both the Earth and the pulsar terms since neglecting the pulsar term has been shown to lead to a biased recovery and lower significance for a present CW \citep{ferranti+2025sourceconfusion}. Additionally, for all CW searches, we include a GWB model through the 14th Fourier mode ($\sim28$ nHz) since \citet{ng15gwb} and this work find evidence for a GWB. For CW searches at low GW frequencies, we either search for a CW in addition to an HD-correlated background or search for it with CURN and resample to an HD-correlated background since \citet{ferranti+2025sourceconfusion} showed that a CURN+CW model could misinterpret a GWB as a CW due to the CW spatial correlations.


\section{Results}\label{sec:results}

\begin{figure*}[ht]
    \centering
    \includegraphics[width=\linewidth]{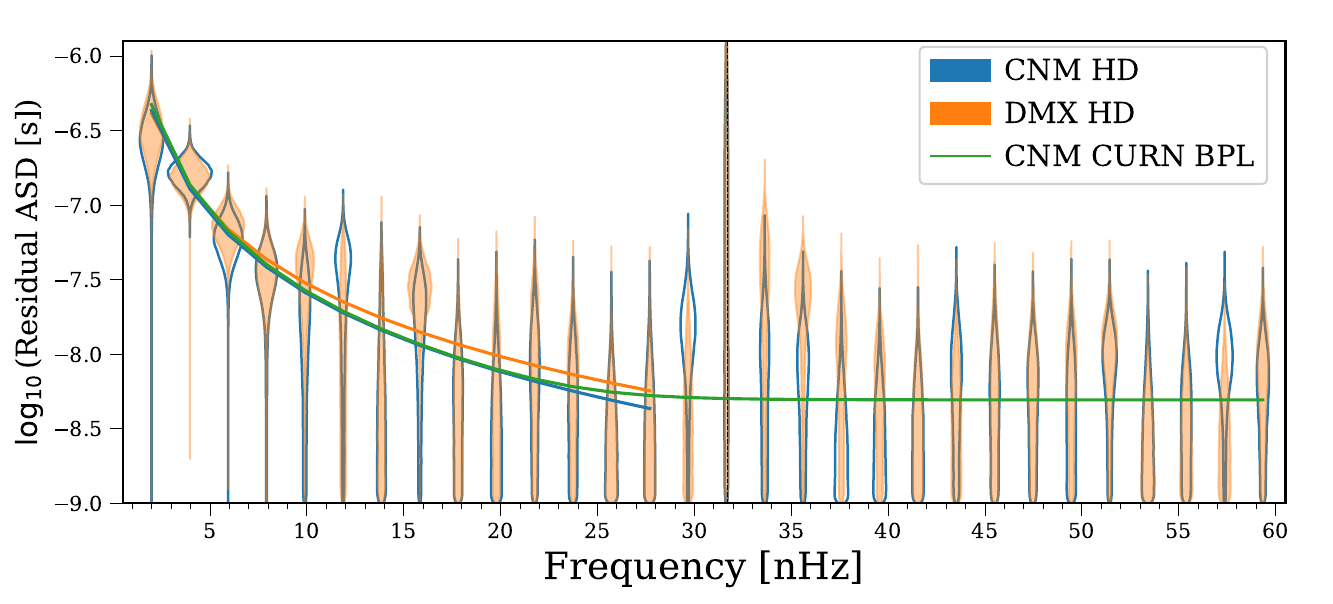}
    
    \vspace{-0.5\baselineskip}
    \caption{\textbf{Changes in the GWB spectrum.} We compare the inference of the HD-correlated GWB model under the different noise models of CNM (blue) and DMX (orange) for different parameterizations of the PSD. The violins compare the full posteriors of a free spectral parameterization while the blue and orange curves compare the MAP power-law spectra modeled through 14 Fourier modes. The green curve is the MAP inference of a broken power law spectrum, which previously has been used to infer the optimal number of Fourier modes, under the CNM. We note that the $16$th Fourier mode is completely unconstrained since it is near the characteristic insensitivity at a frequency of 1/year (denoted by a dashed vertical line) which arises from the marginalization over the linearized timing model with an improper prior.}
    \label{fig:free_spec_comp}
\end{figure*}

Here we reproduce previous GW analyses using the narrowband version of NG15, except we use the CNMs presented in \citet{NG15_CNM} rather than the ``standard'' noise analysis (DMX) which was published in \citet{ng15detchar}. \citet{ng15data} presents the dataset in full detail.

There have been many improvements to pulsar timing array data analysis techniques since the NANOGrav $15$-yr dataset was first released (e.g. \citealt{Laal+2025_solving_PTA, Crisostomi+2025, Susobhanan+2025_vela, Gundersen+2025, vanHaasteren+2025_NP, Curylo+2026}). However, we largely restrict ourselves to the GW models and analysis methods used in \citet{ng15detchar} in order to isolate the impacts of the noise models, rather than the confounding effects of changing multiple components of the data analysis pipeline at once. Combining detailed noise models with advances in pulsar timing data analysis, as well as extended pulsar data sets, will optimize GW sensitivity beyond what we demonstrate in this work.

Our results section presents the new spectral characterizations with CNMs in \S\ref{sec:spectral_characterization}, new significance for HD correlations in \S\ref{sec:detection_statistics}, results of a search for an alternative polarization of gravity in \S\ref{sec:ST_modes}, results from CW searches in \S\ref{sec:cw_searches}, and updates to detector characterization with CNMs in \S\ref{sec:detector_characterization}. 

\begin{figure}[ht]
    \centering
    \includegraphics[width=0.45\textwidth]{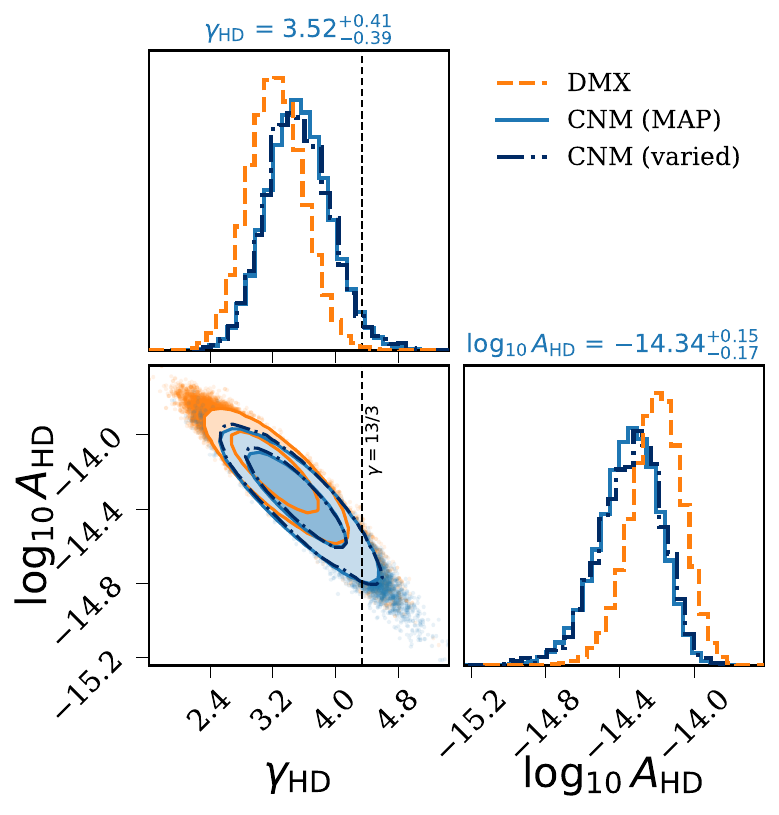}
    \caption{\textbf{Changes in the power law GWB inference.}  Parameter posteriors for a 14 frequency power-law HD model are shown for DMX in orange and CNM in blue. CNM recovers a steeper spectral index, closer to the theoretical value of $\gamma=13/3$ for GW-driven SMBHB mergers (dashed, black line). GWB posteriors under the CNM are nearly identical whether varying the chromatic parameters (dark blue, dot-dashed) or holding them fixed to their single-pulsar MAP values (solid blue).}
    \label{fig:curn_corner_comp}
\end{figure}

\subsection{GWB Spectral Characterization}
\label{sec:spectral_characterization}

The violins in Fig.~\ref{fig:free_spec_comp} show the posterior probabilities for the power spectral density at a number of fixed frequencies, i.e. the $\rho$ parameters from the so-called ``free spectral'' analysis (Eq.~\ref{eq:free_spectral_prior}), effectively a Bayesian periodogram \citep{ng15detchar}. The figure directly compares the inferred GWB spectra between DMX and CNM (assuming chromatic parameters fixed to their MAP values). We note that the 2nd Fourier mode ($f \approx 4$ nHz), which is the most significant in either data set, is more significant with CNM than DMX. We also observe increased power in the 6th, 15th and 29th Fourier modes ($f \approx 12$, $30$, and $58$ nHz) using the CNMs; however the power at these frequencies remains statistically insignificant, with Savage Dickey BFs $<1$ for each Fourier mode. The CNMs noticeably reduce power across frequencies in the $f \approx 31$ to $40$ nHz range. At lower frequencies, we observe reductions in the amplitude and significance of correlated power in the 1st, 3rd, 5th, and 8th Fourier modes ($f \approx 2$, $6$, $10$, and $16$ nHz), where the reduction at 16 nHz is the most substantial change. This frequency was previously flagged as a significant excursion from the PL expectations of the nHz GWB \citep{ng15-discreteness}; we now know this unexpected excursion was a noise-model-dependent effect (as opposed to e.g., a CW \citealt{Goncharov+2026}). We believe the excursion is specifically linked to the modeling of non-stationary chromatic transients in PSR J1713+0747, based on previous achromatic free spectral analyses for this pulsar \citep{Larsen2024, NG15_CNM}.

The HD-correlated free spectrum is the basis for further parameterized spectral modeling \citep{lamb2023rapid}, which we carry out to assess how inferences on the underlying origins of the GWB are impacted by CNMs in \S\ref{sec:astro_cosmo_interp}. The next subsections summarize our results using the more generic PL and BPL spectral models as fit to the timing residuals.

\begin{figure}
    \centering
    \includegraphics[width=\linewidth]{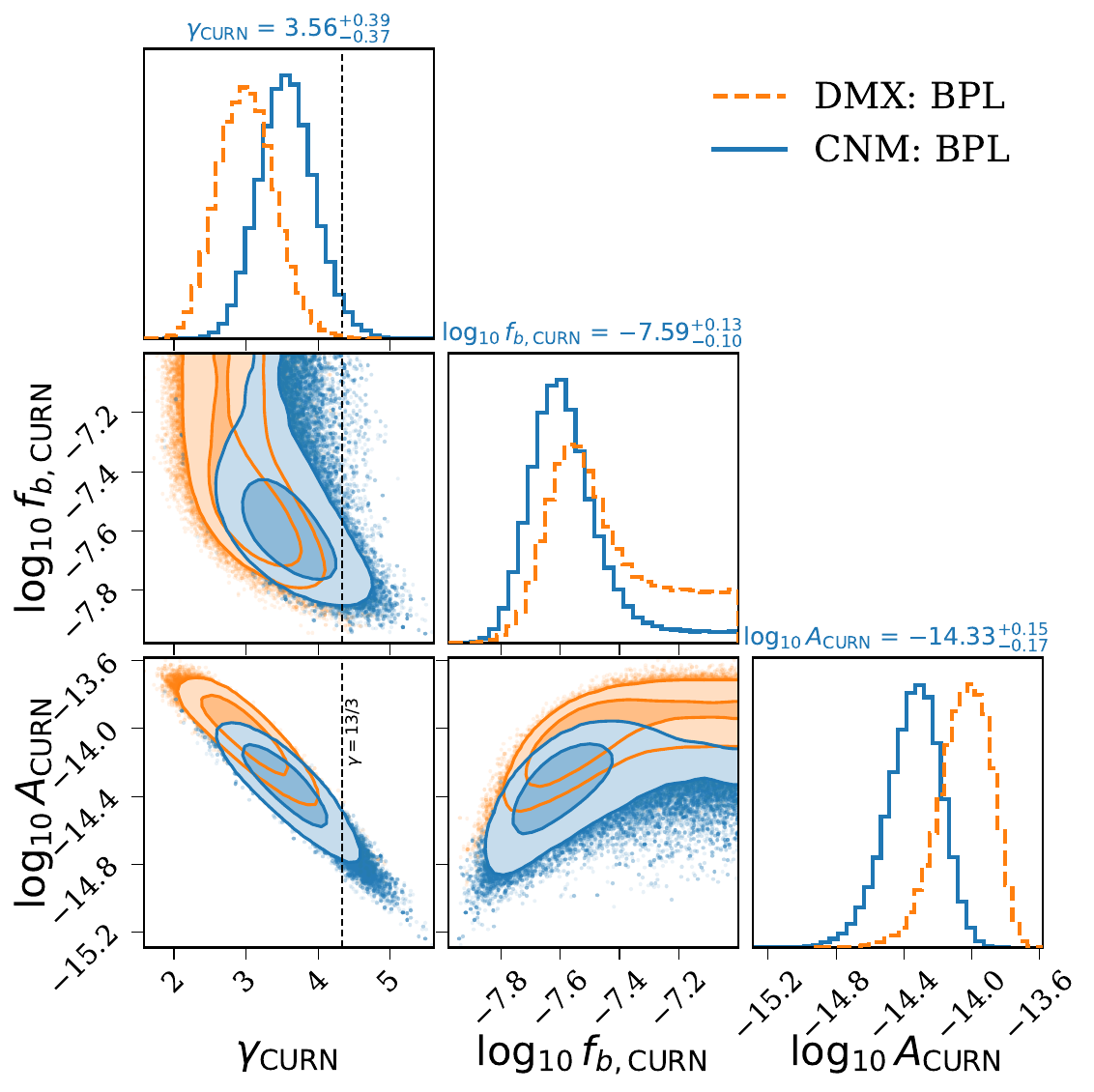}
    \caption{\textbf{Changes in the broken power law GWB inference.} Posteriors of the broken power law CURN parameters from a 30 frequency fit. Orange shows the \texttt{ceffyl} refit of the NG15 DMX run \citep{ng15gwb}, and blue is the CNM (this work). Both fits adopt the broken power-law priors shown in Table~\ref{tab:priors}, with $\ell = 0.1$ held fixed.}
    \label{fig:bpl30$f$}
\end{figure}

\subsubsection{Varied $\gamma$ and Broken Power Law Analyses}

Fig.~\ref{fig:curn_corner_comp} shows our comparison of the posterior hyperparameters $A$ and $\gamma$ of the $14f$ PL spectral model for the GWB (Eq.~\ref{eq:powerlaw_crn}; \citealt{ng15gwb}). When assuming a CURN, we recover a lower amplitude $\log_{10}A_{\rm CURN} = -14.37^{+0.23}_{-0.29}$ and higher spectral index $\gamma_{\rm CURN} = 3.63^{+0.65}_{-0.59}$ (90\% credible). This constitutes roughly an $8\%$ increase in median spectral index $\gamma$ and $33\%$ reduction in median GWB amplitude $A$ from the previous results \citep{ng15gwb}, although the majority of the amplitude decrease is a result of the covariance between $A$ and $\gamma$ resulting from the (arbitrary) choice to anchor the spectral amplitude at $f_{\rm yr} = 1/{\rm yr}$ in Eq.~\ref{eq:powerlaw_crn}. The value of $\gamma = 13/3$ expected for a population of SMBHBs is more consistent with our updated inference, within the 95\% credible region of the posterior. Reweighting the CURN samples to HD (shown in Fig.~\ref{fig:curn_corner_comp}) slightly shifts and broadens the distribution to $\log_{10}A_{\rm HD} = -14.33^{+0.24}_{-0.31}$ and $\gamma_{\rm HD} = 3.51^{+0.72}_{-0.62}$.

These posteriors are estimated assuming \emph{varied} chromatic parameters of the CNM. Holding the chromatic parameters fixed to their MAP values from the single pulsar analysis has nearly no impact on the inference, as shown in Fig.~\ref{fig:curn_corner_comp}. In other words, the NG15 GWB spectral inference is data-dominated rather than prior-dominated with respect to the CNM parameters.

\begin{figure}
    \centering
    \includegraphics[width=\linewidth]{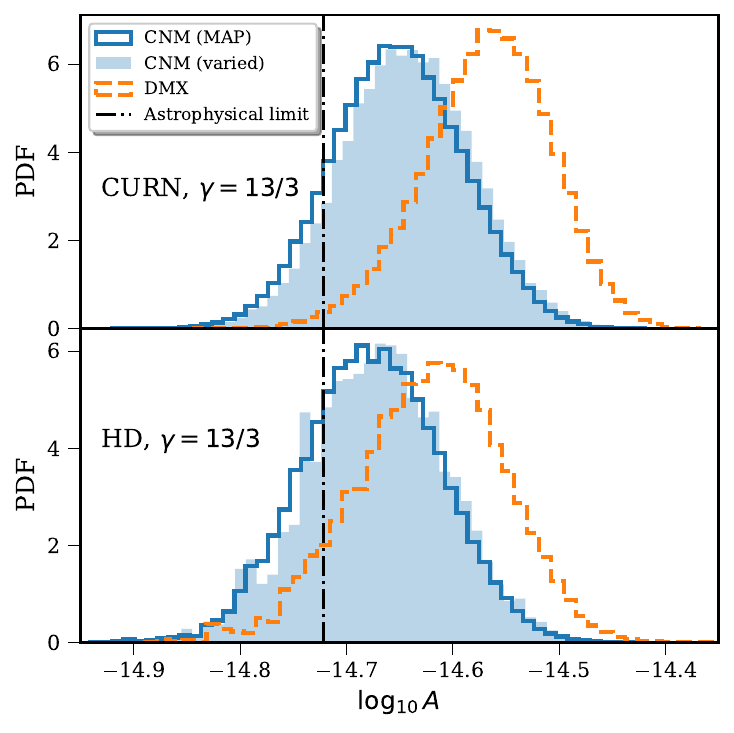}
    
    \vspace{-0.5\baselineskip}
    \caption{\textbf{Changes to the fixed-$\gamma$ GWB inference.} Estimates of the GWB characteristic strain spectral amplitude $A$ at a reference frequency of 1/yr, assuming a 14$f$ power law spectrum with fixed $\gamma = 13/3$ from 2 to 28 nHz, with (top panel) no interpulsar correlations (CURN), and (bottom panel) Hellings-Downs (HD) correlations. We estimate a systematically lower strain amplitude of $A_{\rm HD} = 2.1^{+0.6}_{-0.5}\times 10^{-15}$ using the CNMs (blue) vs the results from \citet{ng15gwb} using the standard DMX chromatic model (orange dashed). Within sampling error, fixing the chromatic parameters to their MAP values during the analysis (blue outlined) does not significantly impact the results over an analysis with varied chromatic parameters (blue shaded). For comparison, we also show a predicted astrophysical limit on the GWB strain amplitude assuming an SMBHB origin, $A \lesssim 1.9\times 10^{-15}$ (dashed black line; \citealt{Mingarelli2026}).} 
    \label{fig:A13/3}
    \vspace{-0.5\baselineskip}
\end{figure}

Despite using a more finely tuned chromatic model following the analyses in \citet{NG15_CNM}, the overall impact on the spectral characterization under the CNM is overall consistent with the ``DMGP'' analysis presented in \citet{ng15gwb}, which also modeled PSR J1713+0747's nonstationary chromatic events. Nonetheless, while \citet{ng15gwb} presented ``DMGP'' simply as an alternative to DMX (which has been the standard in NANOGrav analyses for years), the analyses in \citet{NG15_CNM} confirm that the CNMs do improve overall chromatic characterization of the NG15 timing residual data. As such, we can trust the new spectral characterization with more confidence.

Past CURN analyses have shown the spectral index $\gamma$ may be dependent on the number of Fourier modes included in the model \citep{NANOGrav_12p5yr_gwb, ng15gwb}. The BPL model diagnoses the frequency $f_b$ above which the spectrum becomes flat, and is used typically to set the truncation frequency of the power law model, ensuring the low frequency inference is unbiased by high-frequency systematics. \citet{ng12p5_CNM} found in the case of the NANOGrav 12.5 yr dataset, that implementation of CNMs successfully ameliorated the band-dependence of the spectral inference, corresponding to no significant presence of a tail using the BPL.

For NG15, Fig.~\ref{fig:bpl30$f$} compares the posteriors over $\log_{10}A$, $\gamma$, and $\log_{10}f_b$ under both noise models, assuming a $30f$ BPL spectrum for the CURN (Eq.~\ref{eq:bpl}). Contrary to \citet{ng12p5_CNM}, we find some continued support for the presence of a flat tail even under the CNMs. Furthermore, we find a slightly reduced median value of $f_{b,\rm CURN} = 25.7$ nHz under the CNMs, which is close to the 13th Fourier mode. However, Fig.~\ref{fig:bpl30$f$} also shows noticeable covariances with the $A$ and $\gamma$ parameters, which are substantially different under the CNMs; this covariance is in part responsible for the changing inference on $f_b$ (i.e., for a fixed amplitude of the high frequency tail, the bend frequency must decrease as spectral index increases). Overall, the BPL analysis indicates there is still some presence of non-PL signal at higher frequencies, which could be due to continued presence of unmodeled noise, or perhaps the high-frequency breakdown of the stochasticity of the GWB \citep{ng15-discreteness, Lamb+2025}.

The MAP BPL spectrum is shown in comparison to the PL and free spectra in Fig.~\ref{fig:free_spec_comp}.

\subsubsection{Fixed $\gamma = 13/3$}
\label{sec:fixgam}

An analysis with fixed $\gamma = 13/3$ most directly probes the GWB amplitude $A$ under the assumption of a SMBHB hypothesis in the analytic limit \citep{Phinney2001}. To keep consistent with \citet{ng15gwb}, we assume the GWB model is HD correlated, and spans the same 14 Fourier modes as selected by \citet{ng15gwb} via a broken power-law analysis. Using our CNMs, we estimate the GWB amplitude (median and 90\% credible region) at the reference frequency of 1/yr as $A_{\rm HD}|_{\gamma=13/3} = 2.1^{+0.6}_{-0.5}\times 10^{-15}$. This median estimate is a 12\% reduction from the previously estimated amplitude $A_{\rm HD}|_{\gamma=13/3} = 2.4^{+0.7}_{-0.6}\times 10^{-15}$ by \citet{ng15gwb}, using the same dataset. Besides the choice of pulsar noise models, the only change in analysis method is the use of likelihood reweighting \citep{Hourihane+2023} to estimate the HD-correlated posterior after an initial MCMC analysis which assumes a CURN model with no interpulsar correlations.

Fig.~\ref{fig:A13/3} further compares the GWB amplitude at fixed $\gamma = 13/3$ under different analysis settings, i.e., whether we model the GWB as a CURN or an HD-correlated process. Both posteriors on $A$ under the DMX model are reproduced from \citet{ng15gwb}. One feature of our custom noise analysis is that the estimated amplitude of a CURN, $A_{\rm CURN}|_{\gamma=13/3} = 2.2^{+0.6}_{-0.5}\times10^{-15}$, is substantially closer to the HD-correlated GWB amplitude than DMX, which yields $A_{\rm CURN}|_{\gamma=13/3} = 2.7^{+0.6}_{-0.6}\times10^{-15}$. As a CURN process may model both GWB and genuinely uncorrelated noise components, these results suggest that our chromatic modeling has reduced some component of quasi-common, uncorrelated noise in the PTA, resulting in more consistent amplitudes between the CURN and HD models.

During all of our full-PTA analyses using the custom chromatic models, we hold the chromatic model parameters fixed to their MAP values from single pulsar analyses, which improves the computational tractability of the analysis. To assess the impact of this choice, in Fig.~\ref{fig:A13/3} we also show the fixed $\gamma=13/3$ GWB amplitude posteriors when all chromatic parameters (excepting the free chromatic indices, $\chi$) are allowed to vary. While the amount of sampling error is larger with varied chromatic parameters, we see no significant impact on the GWB spectral characterization, with only a slight increase to the CURN value $A_{\rm CURN}|_{\gamma=13/3} = 2.3^{+0.6}_{-0.5}\times10^{-15}$ when varying the chromatic parameters. 

\subsection{Hellings-Downs Correlation Analysis}
\label{sec:detection_statistics}

\citet{ng15gwb} reported the same significance for the GWB with DMX and with an uncustomized ``DMGP'' analysis. Similarly, \citet{ng12p5_CNM} reported no change in significance for an HD-correlated GWB between DMX and their CNMs on a subset of pulsars in the NANOGrav $12.5$yr dataset; however, this dataset was not sensitive to correlations in either case. Recent work has suggested that better modeling of chromatic noise can lead increased significance for a GWB \citep{DiMarco2024, Ferranti+2025_bw}. We explore the impacts of CNMs on GWB significance in this section.

\begin{figure}
    \centering
    \includegraphics[width=\linewidth]{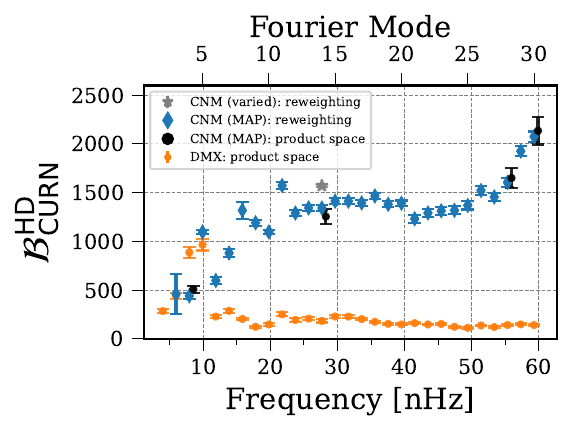}
    \caption{\textbf{HD significance at different frequencies.} The various plot markers denote the BF and uncertainty for HD over CURN modeled out to a specific frequency, with different noise models, and with different BF estimation techniques. The blue diamonds use the reweighting technique with CNM with the chromatic parameters fixed at their MAP values. The black disks use product space sampling and the CNM with the chromatic parameters fixed to their MAP values. The black disks have been slightly offset in frequency for clarity. The small orange disks are BF and uncertainty with DMX using product space sampling. The gray star uses reweighting with CNM but with inference done over all of the chromatic parameters alongside the signal model parameters.}
    \label{fig:bf_vs_f}
    \vspace{\baselineskip}
\end{figure}

\subsubsection{Bayes Factors}
\label{sec:bayes_facs}
\begin{table*}
    \begin{center}
        \begin{tabular}{cc|c|cc|c|c}
            \hline\hline $\mathcal{M}_1$ & $\mathcal{M}_0$ & $\mathcal{M}$ & $\mathcal{B}^{\mathcal{M}+\mathcal{M}_1}_{\mathcal{M}+\mathcal{M}_0}$ (CNM) & $\mathcal{B}^{\mathcal{M}+\mathcal{M}_1}_{\mathcal{M}+\mathcal{M}_0}$ (DMX) & Method & Ref. \\
            \hline HD & CURN & 5$f$ power law & $1096\pm 20$ & $965\pm 60$ & RW & \S\ref{sec:bayes_facs} \\
            HD & CURN & 14$f$ power law & $1342\pm 27$ & $185\pm 16$ & RW & \S\ref{sec:bayes_facs} \\
            HD & CURN & 14$f$ power law, varied CNM & $1571 \pm 14$ & $ - $ & RW & \S\ref{sec:bayes_facs} \\
            HD & CURN & 30$f$ power law & $2074\pm 53$ & $142\pm 9$ & RW & \S\ref{sec:bayes_facs} \\
            \hline HD+ST & ST & 14$f$ power law & $81.29 \pm 11.99$ & $2.21 \pm 0.07$ & SD & \S\ref{sec:ST_modes} \\
            HD+ST & HD & 14$f$ power law & $0.52 \pm 0.01$ & $0.95 \pm 0.02$ & SD & \S\ref{sec:ST_modes} \\
            HD & ST & 14$f$ power law & $157.53 \pm 23.62$ & $2.33 \pm 0.09$ & SD & \S\ref{sec:ST_modes} \\
            \hline
            CW+HD & HD & All-sky, $4<f_{\rm gw}<4.5$nHz & $1.69\pm0.06$ & $0.25\pm0.18$ & SD & \S\ref{sec:all_sky} \\
            CW+CURN & CURN & All-sky, $4<f_{\rm gw}<4.5$nHz & $277\pm95$ & $3.60\pm0.92$ & SD & \S\ref{sec:all_sky} \\
            CW+HD & HD & Targeted (SDSS J0729+4008) & $0.64 \pm 0.05$ & $3.7 \pm 0.1$ & SD & \S\ref{sec:cw_targeted_searches} \\
            CW+HD & HD & Targeted (SDSS J1536+0441) & $0.72 \pm 0.03$ & $1.91 \pm 0.04$ & SD & \S\ref{sec:cw_targeted_searches} \\
            \hline \hline
        \end{tabular}
        \caption{Bayes factors identifying preference for various GW models ($\mathcal{M}_1$ vs $\mathcal{M}_0$) under different analysis settings $\mathcal{M}$, as compared whether using the CNMs or standard DMX chromatic model for the NG15 data. Method column indicates whether the Bayes Factor was computed using likelihood reweighting (RW) or approximated via the Savage-Dickey density ratio (SD).}
        \label{tab:BFs}
    \end{center}
\end{table*}


\begin{figure*}
    \centering
    \includegraphics[width=\linewidth]{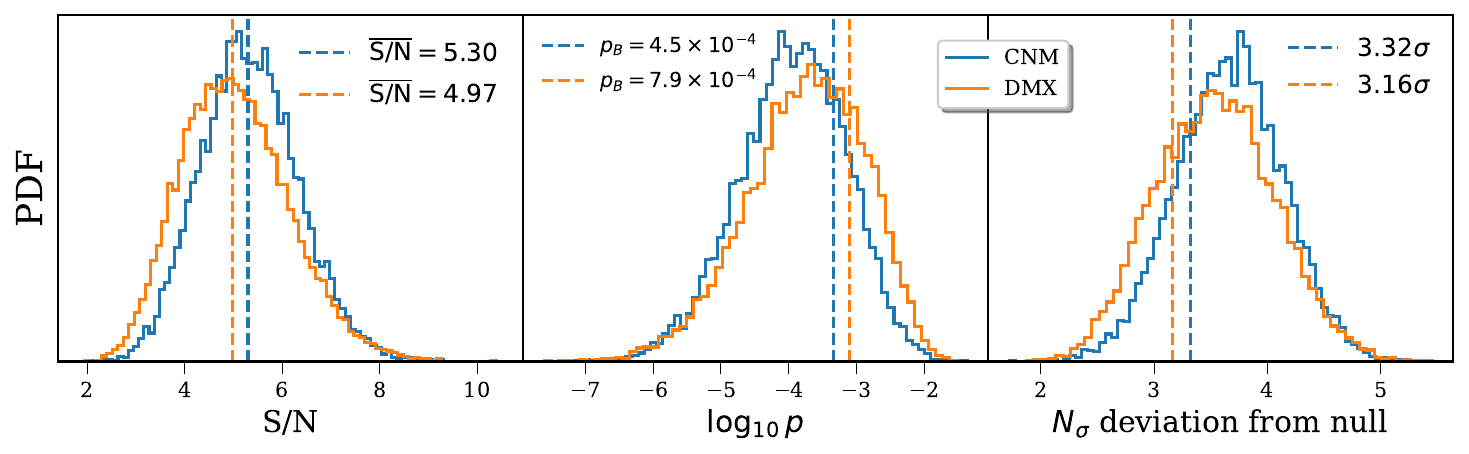}
    \caption{\textbf{Changes in detection statistics for the GWB.} NG15 noise marginalized optimal detection statistics for HD cross-correlations using the CNMs (blue), as compared against reproduced results from \citet{ng15gwb, ng15_ppc} using DMX (orange). \emph{Left:} The distribution of the traditional OS S/N used in \citep{ng15gwb} and given by Eq.~\eqref{eq:OS_SNR}; \emph{Middle}: The distribution of noise marginalized $p$-values computed analytically from the GX2 null distribution as used in \citet{ng15_ppc}; \emph{Right:} The distribution of the $p$-value as mapped to an equivalent number of standard deviations of the normal distribution, computed from the inverse CDF as $N_\sigma = {\rm{CDF}}^{-1}(1 - p)$. The improved noise models yield an improved mean OS S/N from $\overline{\rm S/N} = 5$ to $\overline{\rm S/N} = 5.3$. Meanwhile, $p_B$, computed as the mean of the $p$-value distribution (dashed line), reduces by nearly a factor of 2. These values correspond to a marginal improvement in the $p$-weighted $\sigma$-significance from $3.16\sigma$ to $3.32\sigma$. Each result uses the same 14-frequency, varied-$\gamma$ CURN analysis to obtain the underlying pulsar noise distribution. These analyses complement our Bayesian results to show that improved pulsar noise models yield improved sensitivity to HD cross-correlations.}
    \label{fig:OS-summary}
\end{figure*}

Table~\ref{tab:BFs} presents our detection statistics results for several key searches reanalyzed in this work. Fig.~\ref{fig:bf_vs_f} compares the significance for HD-correlations over CURN as a function of frequency between DMX and CNM. While both the BF for HD over CURN grows for both CNM and DMX between the 2nd and the 5th Fourier modes, at the 6th Fourier mode, the BF with DMX drops down to $\sim250$ and remains around $\sim200$ through the 30th Fourier mode. On the other hand, with CNM the BF drops slightly at the 6th Fourier mode but continues to grow through the 11th to 12th Fourier mode before flattening out to a value around 1400 until the 25th Fourier mode where it begins to grow to $\sim2000$ at the 29th Fourier mode. Values for these Bayes factors for $5,14$, and $30$ Fourier modes can be found in Table~\ref{tab:BFs}. The exact reason for the sharp drop in significance for HD correlations with DMX at the 6th Fourier mode remains unclear, but myriad studies have shown that unmodeled noise can produce lower detection statistics \citep{DiMarco+2025, ng12p5_CNM}, and changing the chromatic models is the only difference between this work and \citet{ng15gwb} so it is fair to surmise that unmodeled or mismodeled chromatic noise is ruining the significance for HD correlations at the $6$th Fourier mode and higher.

Similarly curious is the increase in BF in CNM at high frequencies. It is interesting to note the slight bump in HD-correlated residual ASD at the 29th Fourier mode in Fig.~\ref{fig:free_spec_comp} although this feature in the HD spectrum is not statistically significant. Parameter estimation of a power-law model is biased to be shallower when modeling the GWB out to 30 Fourier modes and slight covariances exist between select pulsar's red noises and this feature in the GWB spectrum. While this increase in BF for HD could imply some high frequency GW source such as an unresolved SMBHB, \S\ref{sec:all_sky} finds no significance for a CW in an all-sky search near these frequencies. Future datasets will clarify the significance and origin of these features in the GWB.

\begin{figure}
    \centering
    \includegraphics[width=\linewidth]{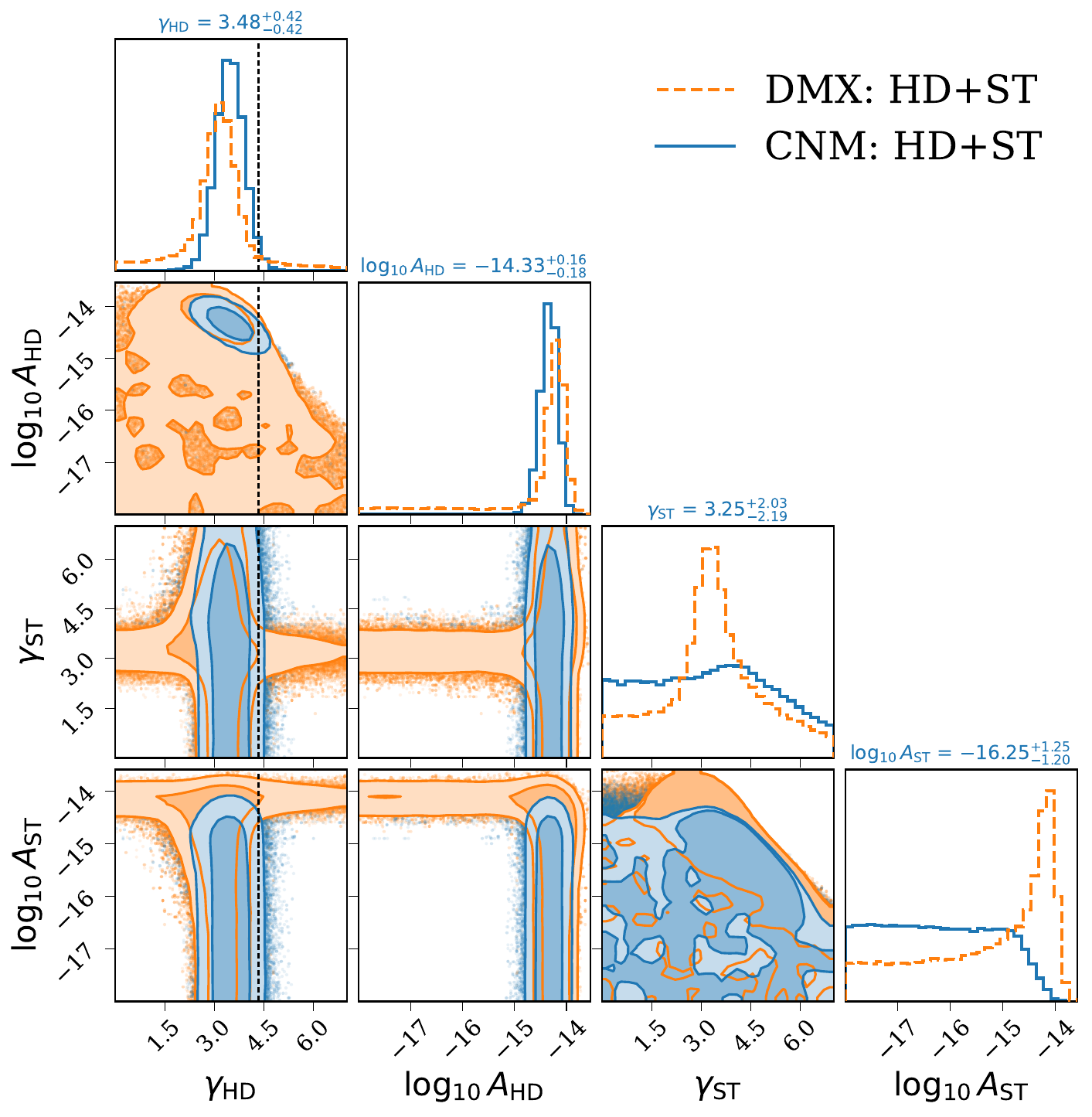}
    \caption{\textbf{Alternative polarization search.} Posteriors for the spectral parameters of a TT-polarized (HD-correlated) GWB and an ST-polarized GWB. Baseline results using the DMX chromatic model (orange) are reproduced from \citet{ng15_altpol}. Using our CNMs (blue) substantially improves the evidence/constraints for the HD-correlated mode, and reduces support for a large ST mode.}
    \label{fig:alt_pol_search}
\end{figure}

\subsubsection{Optimal Statistic}
\label{sec:OS_results}

We next assess how our custom chromatic models impact the evidence for HD correlations in the NG15 dataset using the single-component, noise-marginalized PTA OS. Fig.~\ref{fig:OS-summary} summarizes our results. Firstly, the mean of the S/N distribution improves from $\overline{\rm{S/N}} = 5$ reported in \citet{ng15gwb} using the standard DMX chromatic model to $\overline{\rm{S/N}} = 5.3$ using our custom chromatic models. While this S/N improvement is not as large as the Bayesian analysis would suggest, this still verifies that by improving noise mitigation, our CNMs improve our ability to resolve HD correlations. The low-S/N tail of the distribution is noticeably improved using our custom chromatic models, while the high-S/N tail is practically unchanged. This isolated change in the low-S/N tail could result from the fact that no additional evidence-providing TOAs are included, only improved noise mitigation.

Alongside complementing the Bayes factor, the OS also provides the most computationally expedient method to compute $p$-value detection statistics for the HD curve using the GX2 null distribution. Fig.~\ref{fig:OS-summary} gives the distributions over both values after noise marginalization. After averaging over the distribution we get the Bayesian $p$-value, $p_B$, and corresponding $\sigma$ significance level. \citet{ng15_ppc} first estimated $p_B = 7\times 10^{-4}$ or $3.2\sigma$ using NG15. Our reproduction of this analysis yields slightly higher values $p_B = 7.9 \times 10^{-4}$ (or $3.16\sigma$) using the standard DMX model, but this decreases to $p_B = 4.5 \times 10^{-4}$ (or $3.32\sigma$) using our CNMs. The effects on the high and low S/N tails are also reflected in the $p$-value and $\sigma$ distributions. While custom noise modeling does clearly improve detection significance, higher $\sigma$ is still hard earned. Detailed noise modeling is still secondary to the acquisition of additional high-precision timing data in order to statistically raise the GWB past the traditional 
$5\sigma$ ``discovery'' threshold \citep{IPTA_checklist_2023}.
\newline

\subsection{Search for Scalar Transverse Polarization Modes}
\label{sec:ST_modes}

We have established that our CNMs improve the sensitivity of the NG15 dataset to Hellings-Downs cross-correlations. Naturally, it follows that our noise models may improve sensitivity to any alternative polarization modes of gravity that may be present in nature, beyond the two tensor transverse (TT) polarizations given by general relativity, which determine the shape of the HD curve. Of special interest in the community has been the search for a scalar transverse (ST), or breathing mode, whose contribution is given by the ORF,
\begin{align}
    \label{eq:ST_mode}
    \Gamma^{\rm ST}_{ab} = \frac{1}{8}\left(3 + \cos\zeta_{ab}\right) + \frac{1}{2}\delta_{ab},
\end{align}
a combination of monopolar and dipolar components \citep{ChamberlinSiemens2012}. Past searches for alternative polarization modes have shown these analyses can be highly sensitive to systematics from individual pulsars \citep{abb+21altpol, ng15_altpol}. The solar wind is particularly problematic for these analyses as it can introduce quasi-dipolar correlated signatures in PTAs \citep{tiburzi+2016}. As such, mismodeled solar-wind effects could therefore lend spurious support for a non-Einsteinian polarization mode, particularly the ST mode. Our updated noise models include a state-of-the-art solar-wind model using stochastic and global deterministic components \citep{NG15_CNM}.

Here we reproduce one of the analyses from \citet{ng15_altpol} to assess how our improved pulsar noise models impacts the evidence for the ST polarization mode with NG15. The most physically relevant analysis is a simultaneous search for a ST-polarized GWB component using the ORF of Eq.~\eqref{eq:ST_mode}, on top of an HD-correlated GWB component. Fig.~\ref{fig:alt_pol_search} shows our results using both the DMX model (as reproduced from \citealt{ng15_altpol}) and our CNMs. Using DMX, the full signal could be attributed to either the HD or ST process, with a Bayes Factor $\mathcal{B}^{\rm HD}_{\rm ST} = 2.3 \pm 0.1$ indicating slight but ambiguous support for the HD process, as found by \citet{ng15_altpol}. Using our CNMs, the result is no longer ambiguous; we confidently resolve the HD-correlated component, estimating $\mathcal{B}^{\rm HD}_{\rm ST} \cong 158 \pm 24$ using the Savage-Dickey approximation and the transitivity of Bayes Factors. Further Bayes Factors are given in Table~\ref{tab:BFs}. Meanwhile, the ST component is still unresolved, and the probability of a high-amplitude component is reduced. Assuming a fixed value $\gamma_{\rm ST} = 13/3$ and a uniform prior on $A_{\rm ST}$, the 95\% Bayesian credible upper limit on $A_{\rm ST, 13/3}$ reduces from $(2.24 \pm 0.12) \times 10^{-15}$ to $(1.66 \pm 0.04) \times 10^{-15}$, or a $\sim 26\%$ reduction in ST amplitude at the upper limit.

\citet{ng15_altpol} previously found via dropout analysis that PSRs J0030+0451 and J0613$-$0200 are responsible for a majority of the support for the ST mode in NG15. In \citet{NG15_CNM}, we substantially altered the modeling of both pulsars, which is likely the reason for the reduced support in this work. PSR J0613$-$0200's CNM includes a free-chromatic component for time-variable scattering, which reduces its individual achromatic red noise level at higher GW frequencies. Meanwhile, PSR J0030+0451 is the most sensitive pulsar in the dataset to solar-wind-induced DM variations, for which we have implemented our updated solar-wind model.


\subsection{Continuous Wave Searches}
\label{sec:cw_searches}

In addition to studying how CNMs impact the GWB, we carry out searches for CWs from circular SMBHBs across the entire sky as was done in \citet{ng15singlesource}, and we carry out targeted CW searches with a select set of AGN candidates originally analyzed in \citet{ng15-targeted}. In all searches, we duplicate the signal model exactly from the original search with DMX so that the only difference between the searches is the noise models in use. In all CW searches, we fix the chromatic GP hyperparameters to their MAP values and cache their contributions to the likelihood (cf. \S\ref{sec:cached-chromatic-likelihood}), which alleviates computational challenges and keeps the dimensionality of the search lower. We stress this as a caveat of the CW searches performed here since it is beyond the scope of this work to verify that the chromatic hyperparameters are not covariant with a CW signal. Future work will explore covariances between CW models and other parts of the PTA timing and noise models in detail. At the same time, future CW searches will benefit from next-generation PTA inference suites and improved data analysis techniques \citep{Vallisneri+2025, Laal+2025_solving_PTA, Gundersen+2025}, which will enable inference of chromatic GP hyperparameters alongside CW model parameters with the assistance of likelihood gradient information. In this work, however, we restrict our CW searches to using \texttt{enterprise} and \texttt{PTMCMCSampler} for self-consistency.

\subsubsection{All-Sky Continuous Wave Search}
\label{sec:all_sky}

\begin{figure}
    \centering
    \includegraphics[width=0.48\textwidth]{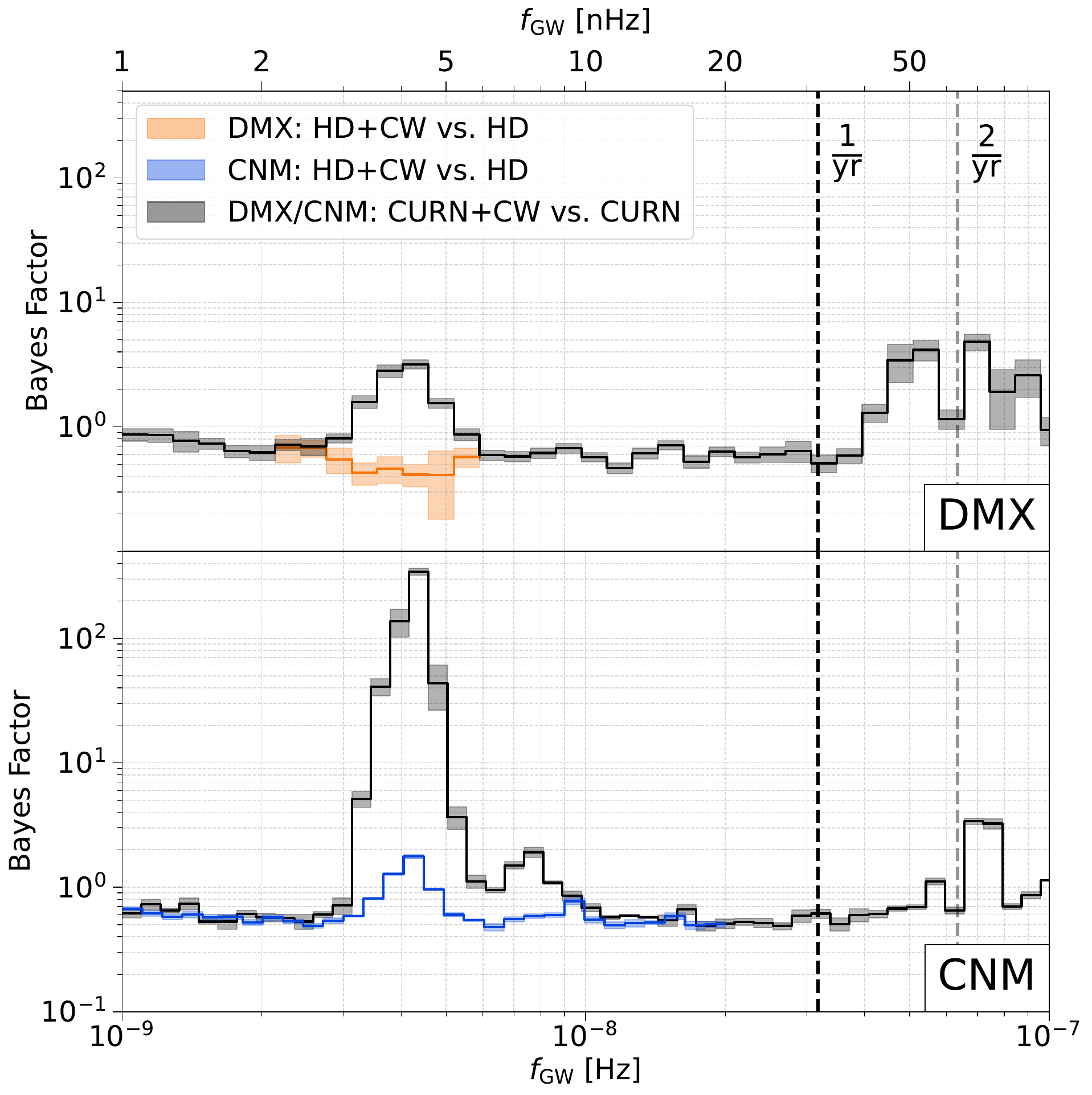}
    \caption{\textbf{All-Sky CW significance} The top panel reproduces a portion of Fig.~1 from \citet{ng15singlesource}. The lower panel updates that analysis with CNM. The black, solid segments with shading represent the BF and its uncertainty for CURN+CW over CURN. The orange and blue segments and shading display the BF for HD+CW over HD with DMX and CNM respectively. Frequencies of 1/yr and 2/yr are plotted with dashed black and gray lines. Note that the more appropriate model comparison for a single source on top of an SMBHB-sourced background is HD+CW vs. HD \citep{ferranti+2025sourceconfusion}, but we reproduce the CURN+CW vs. CURN for comparison.
    } 
    \label{fig:all_sky_BF}
\end{figure}

\begin{figure}
    \centering
    \includegraphics[width=0.49\textwidth]{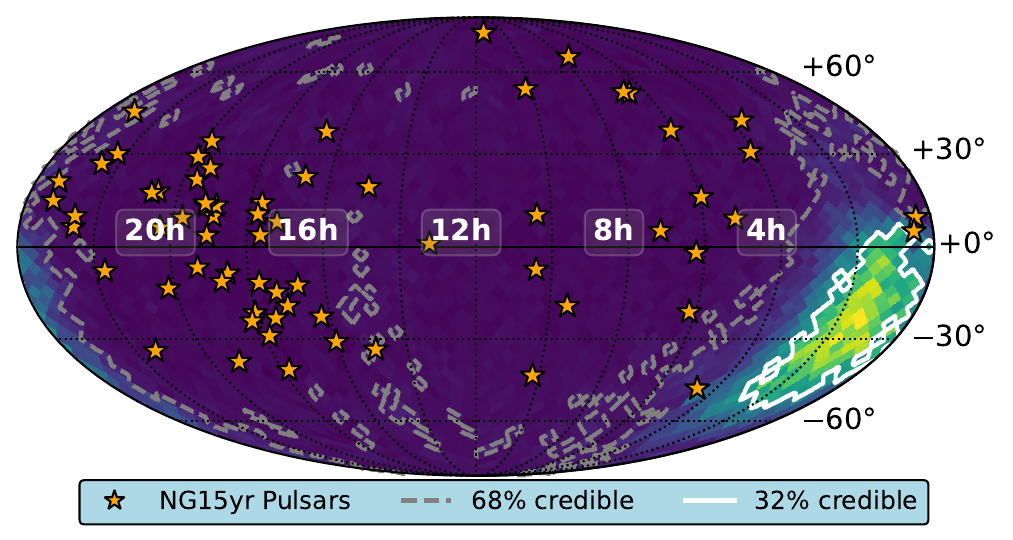}
    \caption{\textbf{Low-frequency candidate sky localization.} 
    The sky map encodes the all-sky posterior for HD+CW for $f_{\rm gw}=4-4.5$nHz. There is no statistical significance for HD+CW over HD in this range (c.f. Table~\ref{tab:BFs}), but the posteriors still display visible structure. The dashed, gray contours enclose $68\%$ and the solid white contours enclose $32\%$ of the posterior density. Beyond the $68\%$ credible region, the sky posteriors are unconstrained across the sky. The $67$ pulsars in the NANOGrav 15-yr dataset are plotted as orange stars.} 
    \label{fig:all_sky_localization}
\end{figure}

We reproduce the results of the all-sky, broad-frequency search, originally carried out in \citet{ng15singlesource} with \texttt{QuickCW} \citep{becsy2022quickCW}. Here, we use \texttt{enterprise} and \texttt{PTMCMCSampler} since \texttt{QuickCW} does not yet natively support the CNMs used in this work.

\citet{ng15singlesource} identified two particular GW frequency bands of interest, one at $\sim4.2$ nHz and one at $\sim170$ nHz, which we will hereafter refer to as the low-frequency candidate and the high-frequency candidate, respectively. Follow up investigations in \citep{ng15singlesource} found that there was significant interplay between the high-frequency candidate CW model and the marginalized, linearized timing model of PSR J$1713+0747$, noting that the candidate lies at a frequency which is suspiciously coincident with PSR J$1713+0747$'s binary orbital frequency. The authors explored using CNMs on a variety of pulsars which appeared to support the high frequency candidate with wide ranging significance for the candidate depending on the number of pulsars with CNMs. The low-frequency candidate, on the other hand, was disfavored after resampling from a CURN+CW model to a HD+CW model.

Fig.~\ref{fig:all_sky_BF} compares the results of the CNM all-sky search performed as part of this work from $1-100$nHz to the all-sky search done with DMX in \citet{ng15singlesource}. We note that we were unable to extend the search beyond $100$nHz to probe the high frequency candidate due to sampling pathologies and computational limitations. The results of the all-sky search near the low-frequency candidate from \citet{ng15singlesource} show a large increase in significance for CURN+CW vs. CURN with CNM as compared to DMX as well as a small increase in support under the HD+CW vs. HD models. We emphasize that CURN+CW is not the astrophysically appropriate model for searching for a CWs at low frequencies given that \citet{ng15gwb} and this work find evidence for an HD-correlated GWB \citep{ferranti+2025sourceconfusion}. We merely include the CURN+CW model as an interesting comparison and note that the BFs and uncertainty presented in Fig.~\ref{fig:all_sky_BF} for CURN+CW vs. CURN are sensitive to the exact frequency binning used in Fig.~\ref{fig:all_sky_BF} since the lower tail of the $\log_{10} h_0$ parameter has few samples in that frequency range. Comparing HD+CW vs. HD, the low-frequency candidate remains statistically insignificant with CNM; we report a BF of $1.69\pm0.06$ for the frequency range $4$nHz$<f_{\rm GW}<4.5$nHz.

Fig.~\ref{fig:all_sky_localization} shows the sky localization of the low-frequency candidate modeled with HD correlations. We again stress that the low frequency CW candidate is not statistically significant when modeled alongside an HD-correlated background. The $32\%$ credible region of sky localization falls in a part of the sky where NANOGrav does not have very many pulsars, but the least sensitive part of the sky at $4$nHz is closer to RA$\approx 06^h00^m$, DEC$\approx 0^\circ00$. \citet{ng15anisotropy} reports no significant evidence for anisotropy on that part of the sky with DMX. \citet{eptadr2_4:single_source} observes a candidate at a similar frequency and presents posteriors which localize it to broadly the same quadrant of the sky, but similar to \citet{ng15singlesource}, they ultimately cannot conclude the nature of the feature in their dataset.

With CNMs, the origins of the low-frequency candidate remain unclear. It seems unlikely that the feature is mismodeled chromatic noise given that candidate's significance increases despite the CNMs more robustly modeling chromatic noise in the 15yr dataset \citep{NG15_CNM}. However, we cannot entirely discount this possibility. The Third Data Release of the IPTA and future NANOGrav data releases may provide more answers to these questions since the significance of an astrophysical source should grow with more pulsars and time span. Future simulations should continue to explore how single sources of GWs might emerge from a GWB, especially in the presence of realistic chromatic noise.

\subsubsection{Targeted Continuous Wave Searches}
\label{sec:cw_targeted_searches}
\citet{ng15-targeted} searched for CWs from $114$ AGN which may host SMBBHs. In this work, we reanalyze $3$ particular AGN using the same signal model (cf. \S\ref{sec:cwmeth}) but with CNM. We reanalyze SDSS J1536+0441 (``Rohan'') and SDSS J0729+4008 (``Gondor'') since \citet{ng15-targeted} reports that their Bayes factors, although only slightly above unity, are outliers among the SMBHB candidates. We also elect to reanalyze 3C66B due to its historic prevalence in PTA CW searches \citep{Iguchi:2010zs, arz:2020}. 

\begin{figure}
    \centering
    \includegraphics[width=0.45\textwidth]{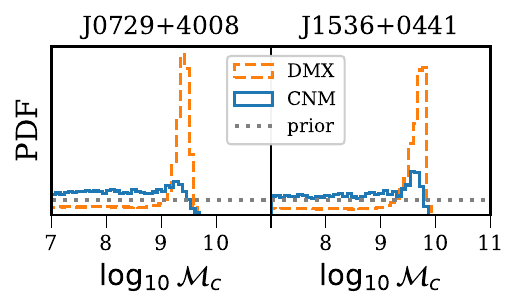}
    \caption{\textbf{Targeted CW searches.} The posteriors for the $\log_{10}\mathcal{M}_c$ parameters are shown for targeted continuous wave candidates J0729+4008 (``Gondor'') and J1536+0441 (``Rohan'') for both the CNM (blue, solid) and with DMX (orange, dashed). For DMX and CNM for both candidates, the posteriors shown are reweighted to include HD-correlations for the GWB. The prior is shown as a dotted gray line. In both cases, the significance for CW candidates are decreased when using custom noise.}
    \label{fig:targeted_searches_comp}
\end{figure}

\begin{figure}[ht]
    \centering
    \includegraphics[width=0.45\textwidth]{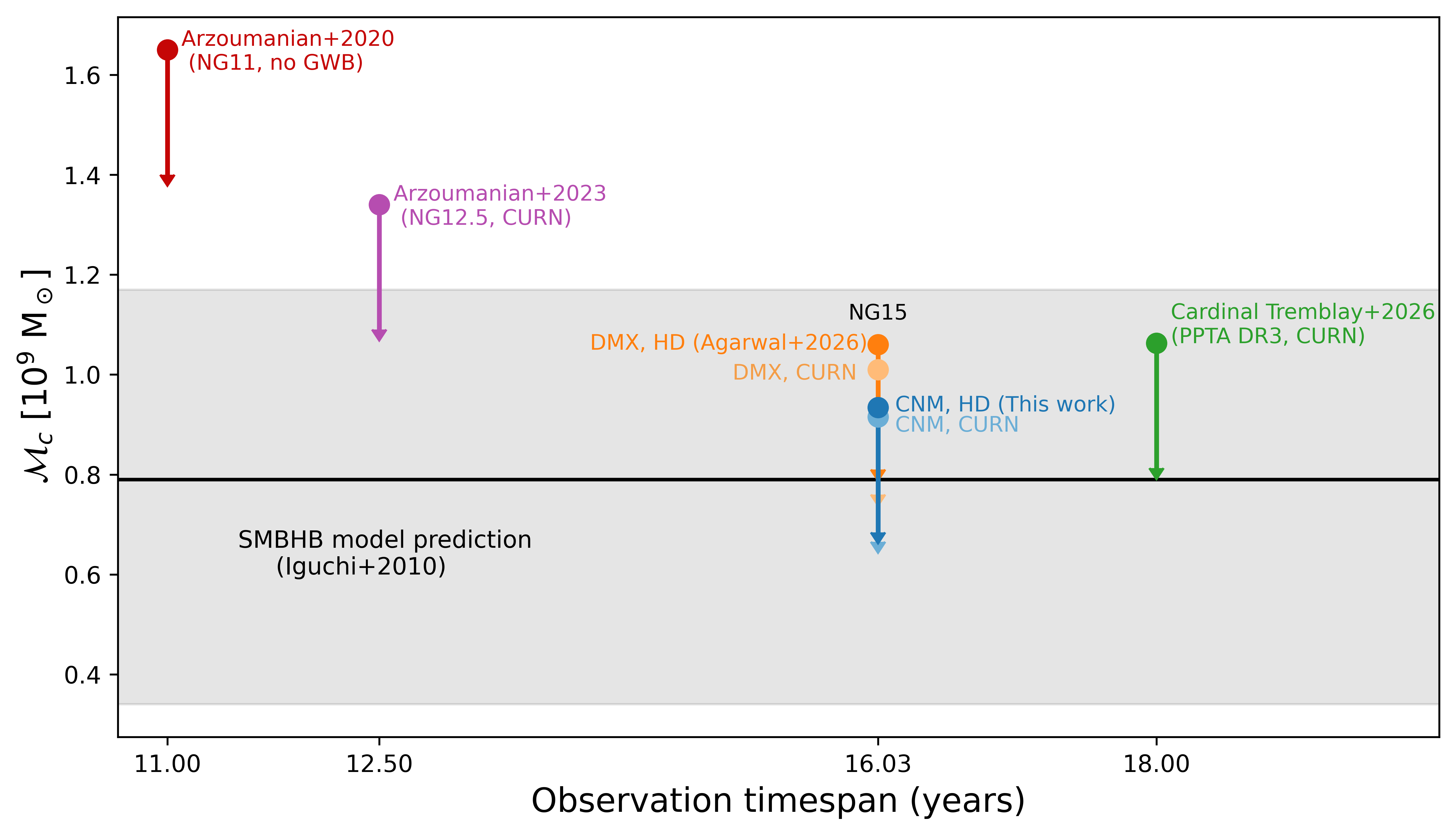}
    \caption{\textbf{Upper limits on the chirp mass for 3C 66B} The solid horizontal line and the grey-shaded region represent the electromagnetically derived constraints on the parameter, from \citet{Iguchi:2010zs}. Most recent NANOGrav \citep{ng15-targeted} and PPTA datasets \citep{Tremblay:2025fgk} probe astrophysically relevant regions of this parameter space. Using CNMs in the 15yr dataset, (which has a total timespan of 16.03 years) the upperlimits decrease to  $\ulcnmcurnthreec$ $M_{\odot}$ with CURN and $\ulcnmhdthreec$ $M_{\odot}$ when including HD correlations. These are the most constraining limits on the parameter using conservative priors to date.}
    \label{fig:3c66b_ul}
\end{figure}

Fig.~\ref{fig:targeted_searches_comp} compares the results from the targeted continuous wave searches for SMBHB candidates J0729+4008 and J1536+0441 with CNM and DMX. Since the luminosity distance is fixed for the targeted searches, the $\mathcal{M}_c$ parameter is used as an amplitude parameter for the Savage Dickey density ratio. Under the CURN + CW model, we report that the significance for both candidates decreases such that a CW model is no longer favored when using CNM. We apply the likelihood reweighting technique \citet{Hourihane+2023} and report that a CW is further disfavored under a HD + CW model with custom noise. See Table~\ref{tab:BFs} for the comparisons of Bayes factors.

\citet{ng15-targeted} ultimately concluded that there was no statistically significant evidence for J0729+4008 or J1536+0441. Random targeting tests further showed that there is nothing notable about the sky locations for targets J0729+4008 and J1536+0441, but narrow GW frequency ranges around $\sim14$ and $\sim21$nHz produce BFs which slightly favor a CW across the sky. A dropout analysis in \citet{ng15-targeted} identified PSRs J0613$-$0200 and J1713+0747 as two of the leading contributors to J0729+4008 and concluded that pulsar-specific noise likely contributed to the support for a CW model. Both \citet{Larsen2024} and \citet{NG15_CNM} found significant changes in the red noise in PSRs J0613$-$0200 and J1713+0747 around $14$nHz. The decreased significance for these candidates with CNMs provides more evidence that CNMs help mitigate noise which is liable to be misinterpreted as low-significance CW signals, which was also found to be true in \citet{Falxa+2023}.

With the NANOGrav 15-year dataset using DMX \citep{ng15-targeted} and the PPTA DR3 \citep{Tremblay:2025fgk}, PTA constraints on the chirp mass of 3C 66B has evolved to probe the EM-based range from \citet{Iguchi:2010zs}. Putative GW emission from 3C 66B at $60.4$ nHz lies near a frequency of reduced GW sensitivity due to the fit for parallax. Nevertheless, with enhanced sensitivity compared to DMX, we revisit this candidate SMBHB with CNMs and constrain its chirp mass to a 95\% Bayesian upper limit of $\ulcnmcurnthreec$ $M_\odot$ compared to $1.01 (1) \times 10^9 M_\odot$ using the DMX models \citep{ng15-targeted} under a CURN hypothesis for the common process. Including HD correlations, the upper limits increase to $\ulcnmhdthreec$ $M_\odot$ with CNMs, as compared to $1.06 (3) \times 10^9 M_{\odot}$ using DMX. With CNMs, we thus rule out the largest region of EM-predicted chirp mass parameter space yet, as seen in Fig.~\ref{fig:3c66b_ul}.

\subsection{Detector Sensitivity \& Red-Noise Budget}
\label{sec:detector_characterization}

\begin{figure}
    \centering
    \includegraphics[width=0.48\textwidth]{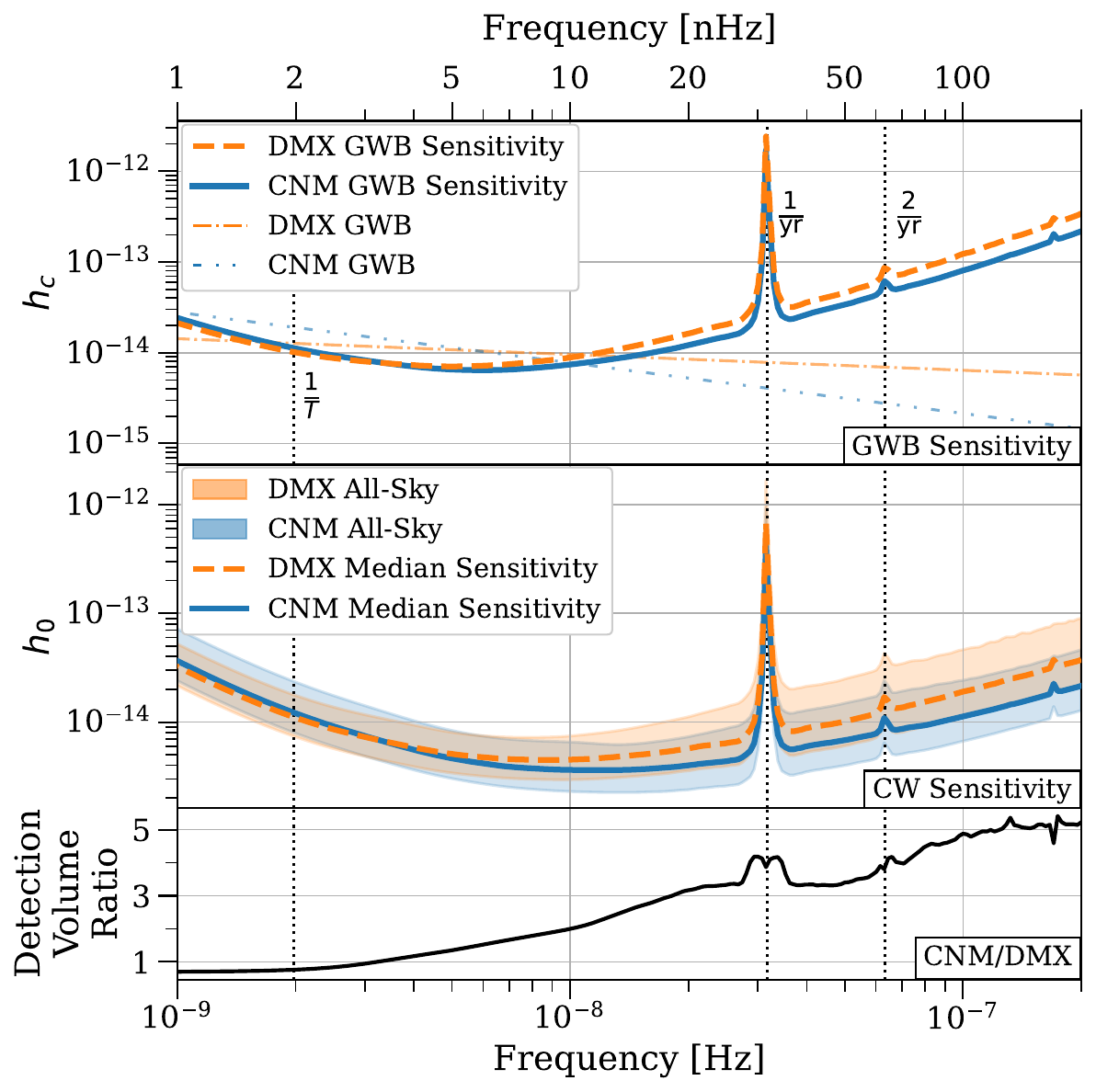}
    \caption{\textbf{Improved GW sensitivity with CNM.}
    The top panel compares DMX (dashed, orange) and CNM (solid, blue) via their GWB sensitivity curves. A power-law GWB is plotted for the MAP values of the background with both DMX and CNM. The middle panel compares the all-sky CW sensitivity by shading a region of sensitivity at each frequency which spans the minimum and maximum sensitivity across the sky at that frequency. The median CW sensitivities are shown via dashed and solid lines for CNM and DMX respectively. Note the units of the top panel are characteristic strain while the middle panel is in strain amplitude, which we convert to by setting a false alarm rate of 0.00135. The bottom panel plots the ratio of CW detection volume with the CNMs to CW detection volume with DMX as a function of frequency.
    }
    \label{fig:sensitivity_comp}
\end{figure}

To assess how these changes in noise models impact the NANOGrav PTA as a detector, we turn to a sensitivity curve formalism. We then update the red-noise budget presented in \citet{NG15_CNM} after the GWB has been removed from the pulsars' intrinsic red noise channel.

\subsubsection{Detector Sensitivity}
\label{sec:sensitivity}
The modern sensitivity curve formalism for PTAs was originally developed in \citet{hazboun:2019sc} and subsequently extended to chromatic GP models in \citet{ipta3p+2024, Gitika+2025}. This formalism combines per-pulsar sensitivity curves into a single PTA-wide sensitivity curve based on frequentist statistics presented in \citet{anholm+2009os, chamberlin+2015os, rosado+2015}. Recently, \citet{baier+2025tuning}, introduced detection volume calculations with sensitivity curves, which we use here but instead of the ratio of the detection volumes between CNM and DMX\footnote{Taking the ratio of detection volumes cancels out the fiducial choices for chirp mass and significance threshold.}. \citet{NG15_CNM} compares single-pulsar sensitivity curves for select pulsars with DMX and the CNM used in this work. It highlights the losses in sensitivity with DMX due to the marginalized timing model as well as an overall reduced white noise floor in the pulsars presented using CNM. We extend this study of the CNM sensitivity by looking at the GWB and CW sensitivity curves for the full PTA, which account for the new spectral characterization of the GWB.

Fig.~\ref{fig:sensitivity_comp} shows DMX and CNM are comparable in sensitivity at at low frequencies, with CNM being only slightly less sensitive while CNM is much more sensitive (i.e. lower strain noise) at mid to higher frequencies in terms of both GWB and CW sensitivity. This can be understood by the fact that we infer a steeper GWB with CNM and the GWB is the largest source of ``self-noise'' at low frequencies since it is composed of both a correlated Earth-term and an uncorrelated pulsar-term \citep{ng15detchar}. \citet{baier+2025tuning} explored the impact of different GWB spectra on detector sensitivity using simulations and found that a steeper power-law GWB shifts the most sensitive frequency higher. We indeed see this here, as the most sensitive frequencies on the CNM CW sensitivity curves (across the sky) fall around $6.5$ nHz while the most sensitive frequencies with DMX lie around $5$ nHz. At high frequencies, where white noise dominates, we see better sensitivity with CNM than with DMX. Interestingly, we find that the sensitivity gains at high frequencies are largest at the sky locations of lowest sensitivity. This can be explained by the fact that regions on the sky of lowest sensitivity have fewer pulsars nearby and so a change in noise in one pulsar in that region has a more dramatic impact on the overall sensitivity in that region. The bottom panel of Fig.~\ref{fig:sensitivity_comp} plots the ratio of CW detection volume for CNM over DMX as a function of frequency, which is slightly less than one at low frequencies but increases to five at high frequencies. Summing these per-frequency detection volumes over an orthogonal set of frequencies which span the all-sky CW frequency range (\S~\ref{sec:all_sky}) gives a total detection volume ratio of $3.2$ for CNM over DMX, which has the straightforward interpretation of being able to probe $3.2\times$ as much of the nearby Universe for SMBHBs with CNMs as compared to DMX.

Future work should explore strategies for marginalizing sensitivity curves over GWB uncertainties as well as exploring the impacts of power-law deviations on GWB sensitivity.

\subsubsection{Red-Noise Budget}
\label{sec:red_noise_populations_analysis}

\begin{figure}[ht]
    \centering
    \includegraphics[width=0.45\textwidth]{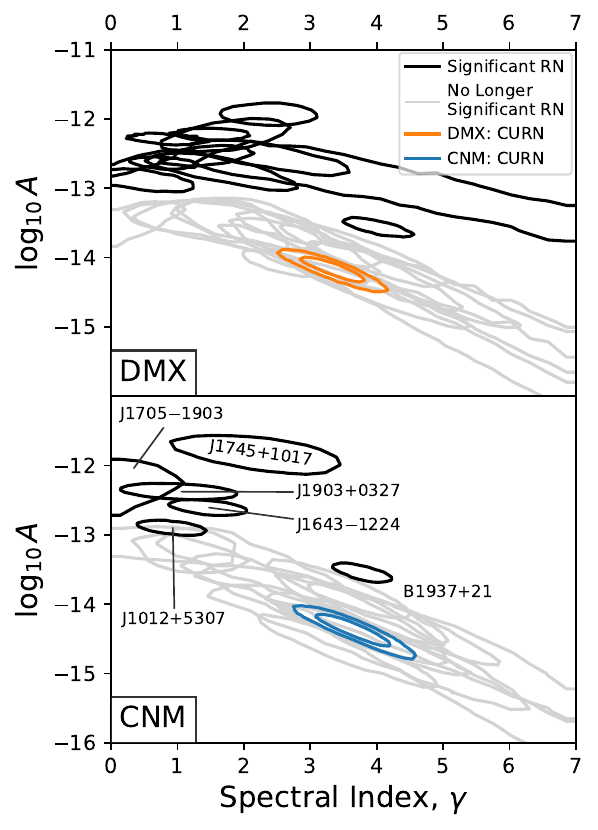}
    \caption{\textbf{Significant red noise populations.} 
    The top panel reproduces Fig.~2 from the NG15 detector characterization paper \citep{ng15detchar} while the bottom panel replicates this analysis with CNM.
    The black contours show the $68\%$ credible regions for pulsars which have significant individual RN when modeled alongside CURN. The grey contours are the $68\%$ credible region of pulsar individual RN which is no longer significant when modeled alongside CURN. The orange and blue contours are the $68\%$ and $95\%$ credible regions for the power-law CURN posteriors with DMX and CNM. The $6$ black contours in the CNM panel are annotated by pulsar name.}
    \label{fig:det_char_fig}
\end{figure}

Intrinsic spin noise in pulsars from instabilities its rotation have been proposed to come from torques in the pulsar magnetosphere \citep{LorimerKramer2004} as well as potential coupling between the neutron star crust and superfluid core \citep{jones1990superfluid}. Since spin noise has been shown to correlate with a pulsar's spin-frequency derivative, millisecond pulsars may have relatively low levels of spin noise compared to canonical pulsars \citep{ lam+2017excessnoise, cordes2026fundamental_noise}. Spin noise in pulsars including milliseconds pulsars remains an active area of research. In this section, we build on the more comprehensive red-noise (RN) budget with CNM presented in \citet{NG15_CNM} by being able to separate the common GWB contributions to a pulsar's achromatic red noise from its intrinsic RN (IRN). We highlight population differences in intrinsic red noise between DMX and CNM. 

Fig.~\ref{fig:det_char_fig} reproduces Fig.~$2$ from \citet{ng15detchar} but updated with the CNM presented in \citet{NG15_CNM}. This analysis reveals which pulsars are contributing to the common process and which maintain significant IRN. \citet{ng15detchar} reported $12$ pulsars which had significant IRN in addition to a CURN process, but with more robust chromatic-noise modeling we report only $6$ pulsars with significant IRN here. Since the achromatic RN models as well as the timing models are identical to those in \citet{ng15detchar}, this implies that what was previously modeled as IRN in these cases is likely mismodeled chromatic noise. We refer to the the population of pulsars traced by gray contours in Fig.~\ref{fig:det_char_fig} as CURN-dominated pulsars since they do not feature any significant IRN during a common process analysis. \citet{ng15detchar} reports $14$ such pulsars, but here we report only $13$, using the same BF criterion of $10^2$ for RN significance. Specifically, PSRs J0023+0923, J1455$-$3330, and J2043+1711 fall into the CURN-dominated population with CNM but are white-noise (WN) dominated with DMX. Both PSRs J1455$-$3330 and J2043+1711 seem to have greatly benefited from their CNM models which led to them becoming CURN-dominated with CNM. Notably, \citet{Lam+2025-J1455} predicted that this would happen for PSR J1455$-$3330 with a simplified DM model. PSR J0023+0923 appears to be an edge case where its single pulsar RN does not look similar to the CURN spectra with either DMX or CNM but it is significance is borderline in both cases. On the other hand, PSRs J0437$-$4715, J1600$-$3053, J1738+0333, and J1744$-$1134 were categorized as CURN-dominated with DMX and are now categorized as WN dominated with CNM. PSR J0437$-$4715,  which is a known scatterer \citep{reardon+2020-J0437}, had a shallow-spectrum RN with DMX, likely contributed to biases in the DMX CURN spectrum as this pulsar showed no significant RN with CNM and had a dropout factor which indicated it was not contributing evidence for the common process in \citet{ng15detchar}. Similarly, PSRs J1600$-$3053, J1738+0333, and J1744$-$3744 also favored inclusion of a free chromatic noise component in \citet{NG15_CNM}, and we see their RN go insignificant with CNMs, further suggesting that these non-dispersive chromatic variations can leak into common noise channels and bias CURN spectral recovery. We note that we do not observe any cases of pulsars which go from being IRN-dominated with DMX to CURN-dominated with CNM.

With CNMs we show that $61$ of the $67$ millisecond pulsars in the NANOGrav 15yr dataset do not show evidence for achromatic RN beyond the stochastic background. Of the $6$ pulsars which retain their IRN, \citet{NG15_CNM} offers several compelling reasons why the majority of these remaining shallow-spectra IRNs may not be from spin noise in the millisecond pulsars. Both PSRs J1705$-$1903 and J1745+1017 are eclipsing black-widow pulsars, which along with PSR J1012+5307 have binary orbital period's $<1$ day. Since these pulsars still have a loud, shallow spectrum red noise both with both DMX and CNM, it is evident that the timing and noise models for these pulsars are still insufficient to capture these complexities. Further, we note that $0<\gamma<1$ has been shown to be unphysical since the covariance function does not converge \citep{vHaastern+2009, vHaastern&Levin2013}. PSRs J1903+0327 and J1643$-$1224 are two of the most chromatically complicated pulsars in the NANOGrav 15yr dataset. PSR J1903+0327 is one of the highest DM pulsars in the dataset and has a stellar mass companion which likely contributes to stellar wind delays which our current models lack a dedicated channel for. PSR J1643$-$1224 lies behind an \textsc{Hii} region in the ISM \citep{Ocker+2024} and both of these pulsars suffer from large scattering delays.
B1937+21's red noise, on the other hand, has a much steeper spectrum, which makes it potentially in line with with predictions for spin noise from magnetospheric torques \citep{melatos+2014} as well as other purposed mechanisms \citep{shannon+2013asteroidbelt}.

As a whole, our GW searches with CNMs prove to be a fruitful endeavor. In Fig.~\ref{fig:free_spec_comp}, we observe shifts in the GWB spectrum which we attribute to better chromatic modeling. These modeling improvements also show substantial gains in significance for HD correlations, especially at frequencies between $10-30$ nHz as well as much greater CW sensitivity at mid to high frequencies due to a reduced white-noise floor. We have shown decreased significance for signals which were previously thought to be spurious such as the ST correlations and specific targeted-CW searches in \citet{ng15-targeted}. Lastly, we show a reduction in the number of significant IRN detections when using CNM. In the next section, we reinterpret our new GW spectrum from the lens of both an astrophysical and cosmological GWB.

\section{Astrophysical and Cosmological Interpretations}
\label{sec:astro_cosmo_interp}

The underlying source or sources of the nanohertz GWB is of major scientific importance. Here we revisit previous inferences on the origins of the GWB using the CNMs. The most common interpretation is that the GWB is created by the stochastic sum of SMBH binaries, i.e., an astrophysical interpretation, which we discuss in section \ref{sec:astro_interp}.  Other explanations involving more exotic explanations such as new particle physics in the early Universe, i.e., a cosmological interpretation, which we discuss in section \ref{sec:cosmo_interp}, are also consistent with the data. 

\subsection{Astrophysical Interpretations}
\label{sec:astro_interp}

\begin{figure*}
    \centering
    \includegraphics[width=\textwidth]
    {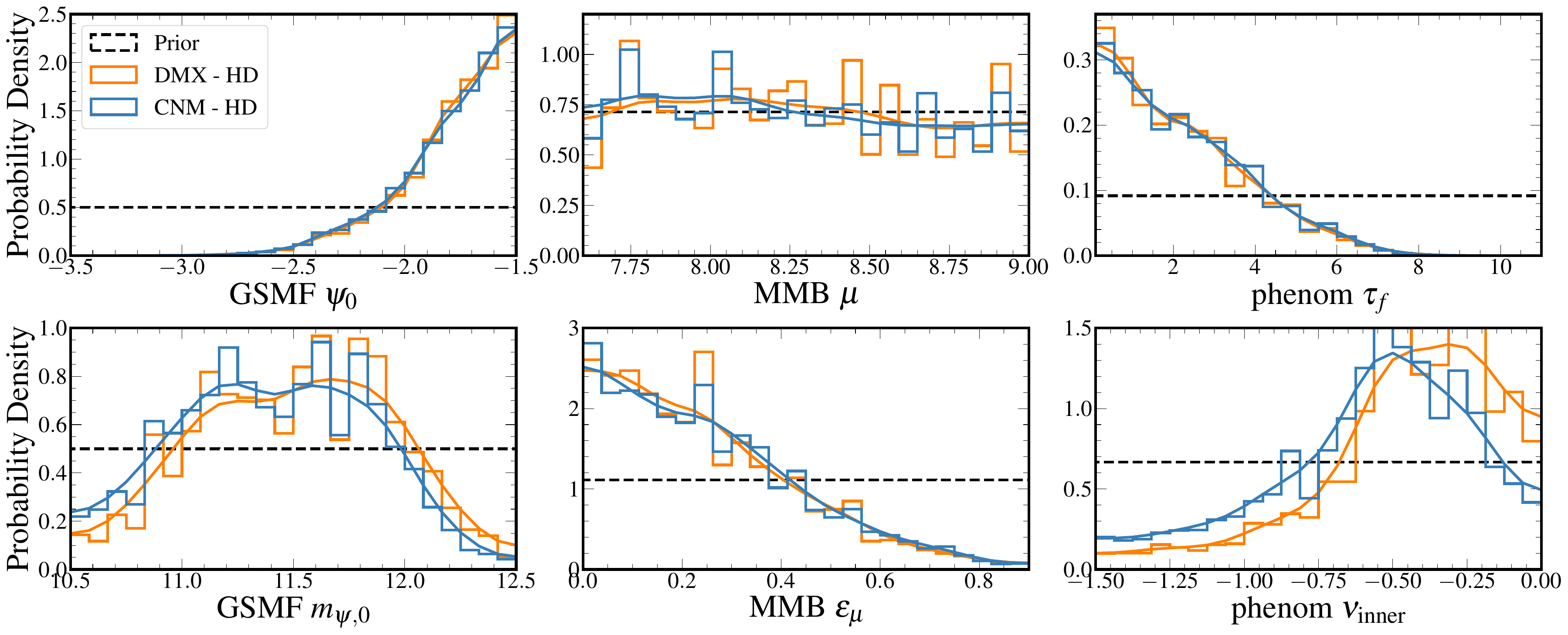}
    \caption{\textbf{Astrophysical population inference.} Marginalized posteriors for the astrophysical parameters fitted against the DMX (orange) and the CNM (blue) HD-only free spectra. Histograms show our raw results and we show smoothed KDEs to aid in interpretation. Priors are shown as black dashed lines and were the same for both fits. The $M_{\rm BH}$--$M_{\rm bulge}$ normalization $\mu$ and intrinsic scatter $\varepsilon_\mu$, as well as the GSMF normalization $\psi_0$ and binary hardening time $\tau_f$ distributions are nearly identical. The best-fit parameters to the CNM spectrum favor slightly lower values of the galaxy characteristic mass, $m_{\psi,0}$, and inner hardening slope, $\nu_{\rm inner}$ ($\sim0.1$ lower in both cases). See \citet{ng15smbbh} for the details of each parameter.} 
    \label{fig:astro_posteriors}
\end{figure*}

\begin{figure}
    \centering
    \includegraphics[width=\columnwidth]
    {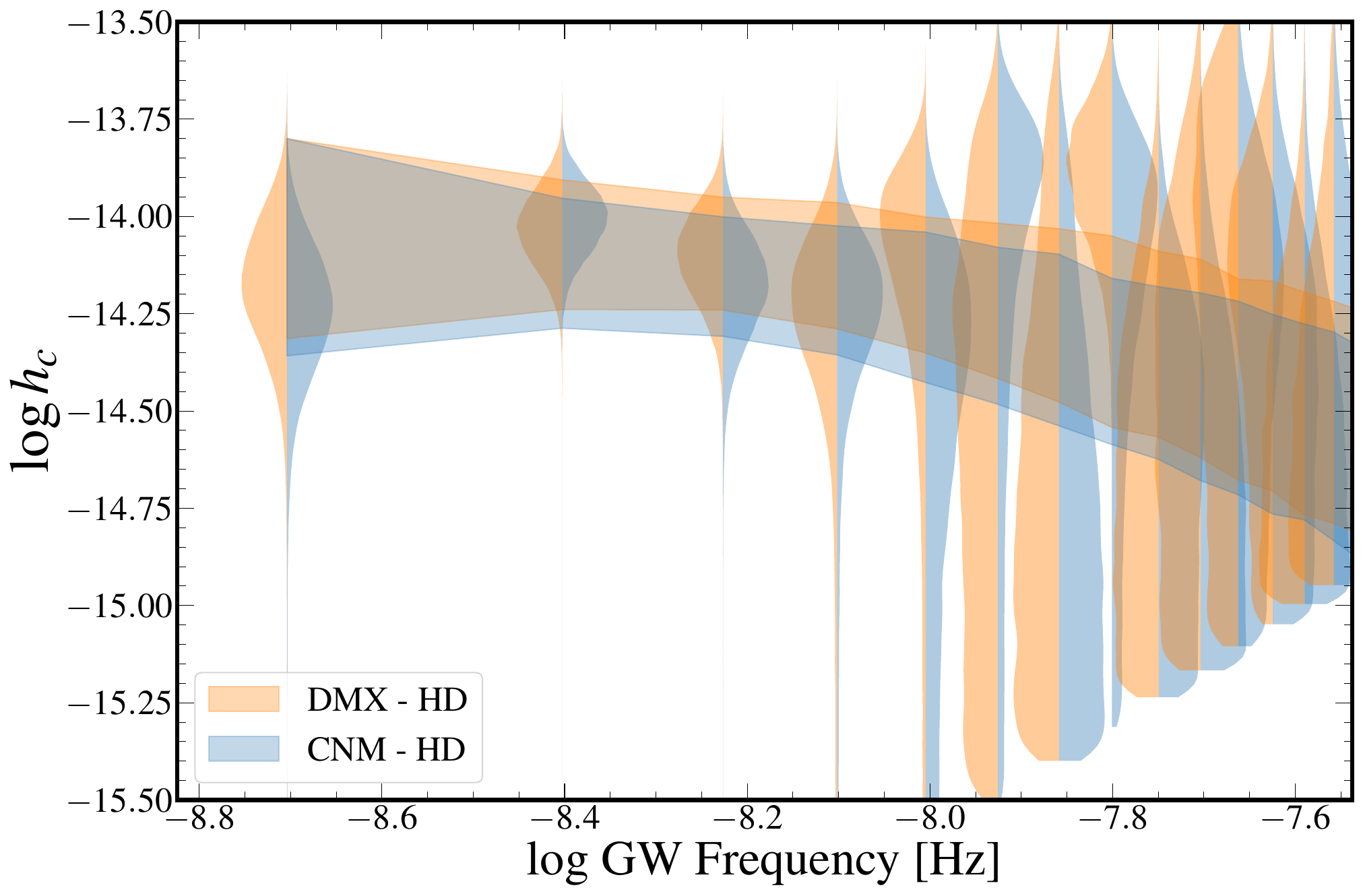}
    \caption{\textbf{Astrophysical spectral fits.} Free-spectral representation of the GWB spectrum with astrophysical fits to the spectra. The violins represent the KDE reconstructions of the DMX (orange) CNM (blue) HD-only free spectrum.  The correspondingly colored shaded regions are the 68\% confidence regions for the best-fit astrophysical spectra to the data for the two noise models.  With the free-spectrum representation of the GWB, we see that some frequency bins increase and others decrease in amplitude between noise models, but the overall best-fit spectrum for CNM is slightly lower in amplitude ($\sim0.06\,\mathrm{dex}$) than that of DMX.} 
    \label{fig:spectra}
\end{figure}

Here we reproduce select analyses from \citet{ng15smbbh} to understand how the CNMs impact inferences of underlying astrophysical populations, assuming an SMBHB origin of the GWB. We used the same \textit{Phenom} model with uniform priors as in \citet{ng15smbbh} (also see Table \ref{tab:priors}). We used models from \texttt{holodeck} to fit to kernel density estimator (KDE) reconstructions of the 14 lowest Fourier modes of the CNM HD-only free-spectrum \citep{lamb2023rapid} using the direct likelihood method described in Appendix C of \citet{ng15smbbh}. The posterior distributions are shown in Fig.\ \ref{fig:astro_posteriors} in blue and are compared to the DMX free-spectrum posteriors in orange. Here the DMX free-spectrum includes the two code corrections detailed in \citet{ng15smbbh_err}. We do not model the monopole, dipole, or CURN components in this analysis, as in \citet{ng15smbbh}, for reasons of computational cost. We see that four of the parameters are nearly identical, but both the galaxy characteristic mass, $m_{\psi,0}$, and the inner hardening slope, $\nu_{\rm inner}$,  now favor slightly lower values.

Recently, \citet{Laal_2025} found that the \textit{Phenom} model we use here is most sensitive to the galaxy stellar mass function (GSMF) and binary hardening parameters, offering a natural explanation as to why the differences we find are limited to these parameters. Because the GSMF characteristic mass dictates where the knee of the Schechter function sits, lower characteristic masses lead to lower number densities of the most massive galaxies. We calculate the SMBH binary population in \texttt{holodeck} using a convolution between the GSMF and the $M_\mathrm{BH}$--$M_\mathrm{bulge}$, such that larger galaxies directly correspond to larger SMBHs and vice versa. Therefore the lower values of $m_{\psi,0}$ indicate lower number densities of the most massive SMBHs, though the change we see here is only slight. Interestingly, the posterior distribution for $\nu_{\rm inner}$ also appears shifted towards lower values. The $\nu_{\rm inner}$ parameter controls the slope of the power law corresponding to the astrophysically driven inspiral of SMBH binaries at separations less than $r_{\rm char}$, which we set to $100 {\rm pc}$ as in \citet{ng15smbbh}. It is therefore a parameterization of the binary hardening driven by processes such as stellar scattering and circumbinary gas disk interactions.
For fixed values of hardening time $\tau_f$, lower values of $\nu_{\rm inner}$ indicate that the astrophysical hardening
phase ceases to dominate at larger separations, such that SMBH pairs spend more time in the GW regime. The result of this is to decrease the low-frequency turnover of the GWB spectrum, i.e., to increase the amplitude at low frequencies.

In Figure \ref{fig:spectra} we show the 68\% confidence regions for the best-fit DMX and CNM spectra (orange and blue) associated with the posterior distributions. The KDE reconstructions of the free spectra for each model are represented by the colored violins. Comparing the violins, we see that the CNM leads to a notable decrease in the characteristic strain amplitude of the fifth lowest frequency bin and a narrowing of the confidence interval for the 6th lowest bin. There is not a uniform change across frequency bins; the amplitude increases in some areas and decreases in others. However, as suggested by the power-law inference discussed previously, there does appear to be a net shift toward lower amplitudes with CNM when we compare the best-fit spectra. The amplitude of the CNM best-fit free spectrum is lower at all frequencies than that of the DMX spectrum with the most notable decrease at the highest frequencies. This is consistent with the changes in the posterior parameters.

Overall, the fits are markedly similar to those of \citet{ng15smbbh}, indicating that our astrophysical interpretation does not change. In particular, the GWB is well explained by a population of SMBH binaries.

\subsection{Cosmological Interpretation}
\label{sec:cosmo_interp}

\begin{figure*}
    \centering
    \includegraphics[width=0.49\linewidth]{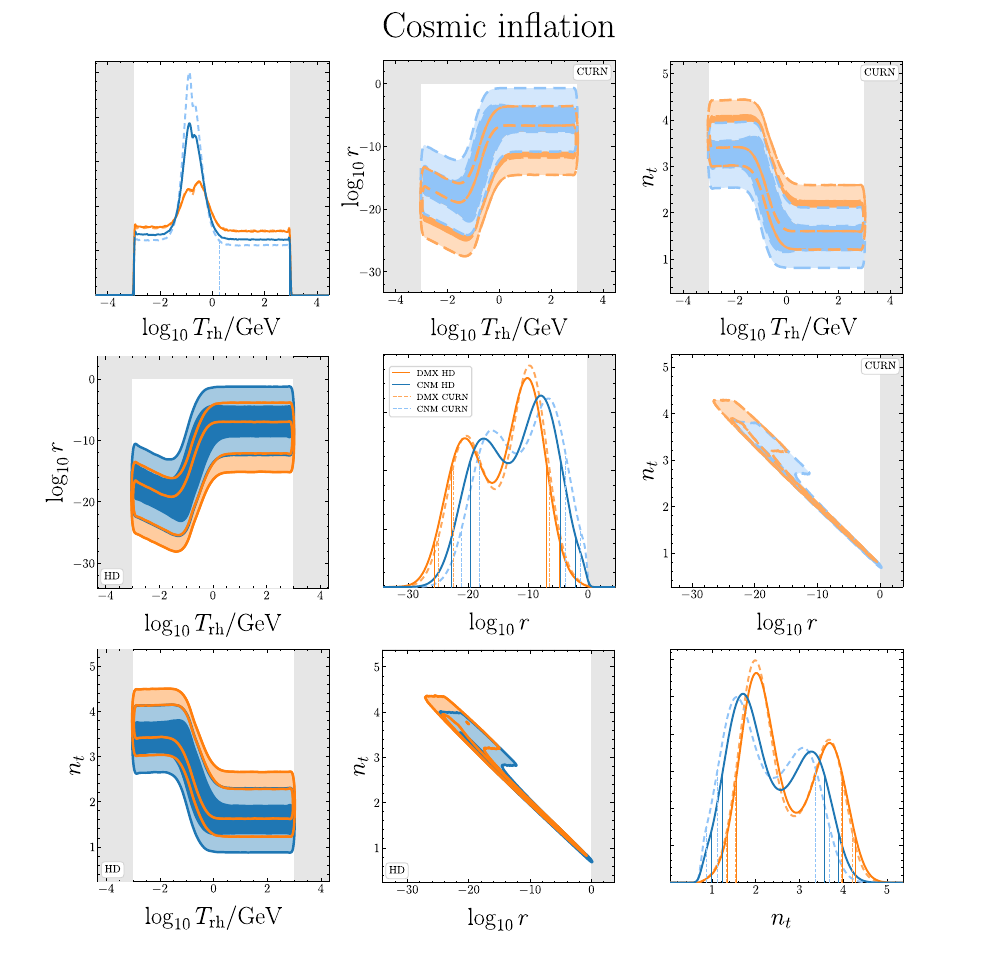}\quad\includegraphics[width=0.49\linewidth]{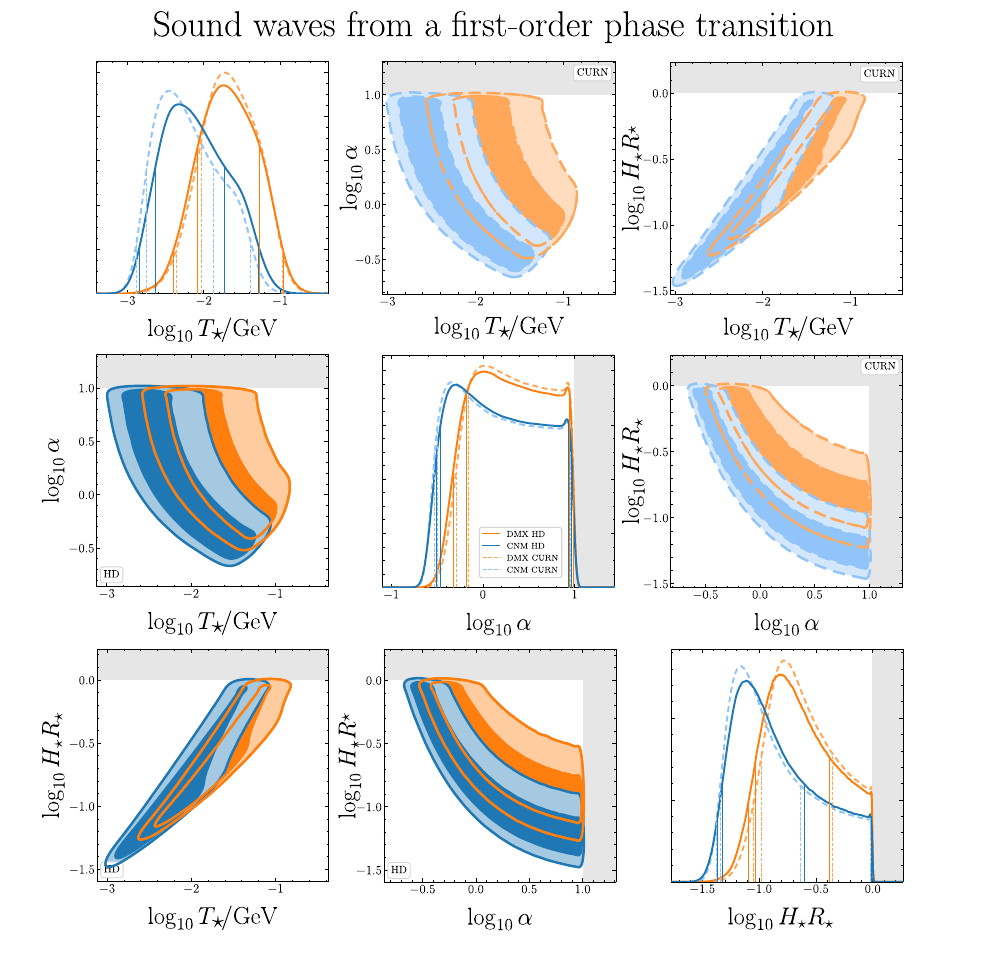}
    \caption{\textbf{Inflation and Cosmological Phase Transition.} ``Double corner plots'' for two exotic interpretations of the NG15 signal: GWs from cosmic inflation (left) and GWs from sound waves produced during a first-order cosmological phase transition (right). We show the results of refits to the NG15 CURN free spectra (plots on the diagonals and upper off-diagonals, dashed lines) as well as to the NG15 HD free spectra (plots on the diagonals and lower off-diagonals, solid lines), for both the old (DMX, orange) and new (CNM, blue) pulsar noise models. The gray-shaded areas mark regions beyond our prior ranges, which coincide with the priors in~\cite{ng15newphysics}.} 
    \label{fig:inflationPT}
\end{figure*}

While SMBHBs are considered to be the leading interpretation of the NG15 signal, other, more exotic explanations\,---\,dynamical processes driven by new particle physics in the early Universe\,---\,remain viable at present. Referring to the SMBHB hypothesis as the astrophysical interpretation, these alternative hypotheses are subsumed under the cosmological interpretation. In~\cite{ng15newphysics}, we investigated the cosmological interpretation of the NG15 signal in detail, fitting an array of early-Universe GWB models to the NG15 data; for a related analysis by EPTA, see \cite{eptadr2_5:gw_sources}. 

\begin{figure}
    \centering
    \includegraphics[width=\linewidth]{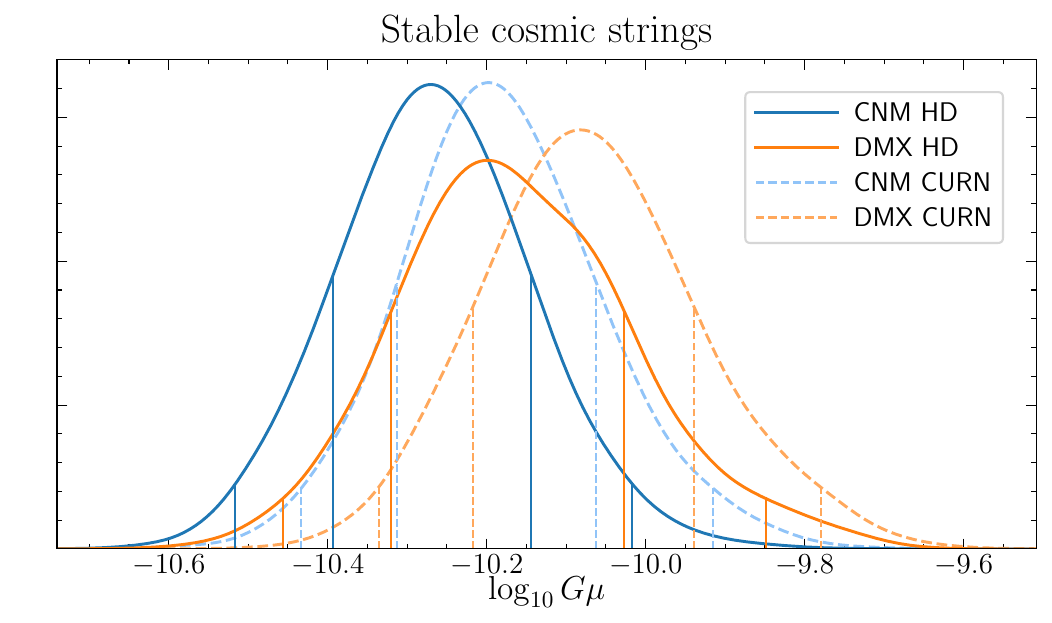}
    \caption{\textbf{Cosmic Strings.} Parameter inference for a third exotic interpretations of the NG15 signal: GWs from stable cosmic strings. As in Fig.~\ref{fig:inflationPT}, dashed and solid lines correspond to refits to the NG15 CURN and HD free spectra, respectively; orange and blue lines distinguish between DMX and CNM.} 
    \label{fig:strings}
\end{figure}

Here, we shall illustrate the effect of our improved pulsar noise models on the parameter inference in select new-physics scenarios that are representative of the broad range of GWB templates studied in~\cite{ng15newphysics}: GWs from cosmic inflation (left panel of Fig.~\ref{fig:inflationPT}), GWs from sound waves produced during a first-order phase transition (right panel of Fig.~\ref{fig:inflationPT}), and GWs from stable cosmic strings in the Nambu--Goto approximation (Fig.~\ref{fig:strings}). In all three cases, we use the same templates for the GWB spectrum as in our 2023 analysis; in the case of the phase-transition model, this notably means that we use the template presented in the erratum \citep{NANOGrav:2023hvmErratum} of \cite{ng15newphysics}. 

In~\cite{ng15newphysics}, we had fitted all exotic spectral models to the NG15 timing residuals, using our \texttt{enterprise} wrapper \ptarcade~\citep{Mitridate:2023oar} in \textit{enterprise} mode. In the present analysis, we resort instead to refits of the NG15 free spectrum (i.e., the NG15 ``violins'' in Fig.~\ref{fig:free_spec_comp}), using \ptarcade in \textit{ceffyl} mode~\citep{lamb2023rapid}. This approach ensures that the DMX and CNM results presented in Figs.~\ref{fig:inflationPT} and \ref{fig:strings} are on the same footing, enabling us to directly compare them to each other. In future work, we plan to make CNM analyses available through \ptarcade as well. 

We refit each new-physics model to the CURN and HD versions of the NG15 free spectrum, specifically, to the first 14 free-spectrum amplitudes (violins) in accord with our analysis in~\cite{ng15newphysics}. The overall conclusion from our refit analysis is that the new pulsar noise models lead to slight quantitative shifts in our 1D and 2D posterior distributions, but to no drastic qualitative changes. All distributions more or less retain their rough shape and are just mildly shifted.

The shifts that we find are consistent with the power-law reconstruction of the signal in Fig.~\ref{fig:curn_corner_comp}. The CNM analysis of the data prefers slightly larger values of the spectral index $\gamma$, which translates to a slightly smaller tilt of the GW energy density power spectrum $\Omega_{\rm GW}$. This is precisely what we observe for the tensor spectral index $n_t$ in Fig.~\ref{fig:inflationPT}, whose posterior distribution is now shifted to slightly smaller values. At the same time, because of the strong anti-correlation between $n_t$ and the tensor-to-scalar ratio $r$, the posterior distribution for $r$ is shifted to larger values. The reheating temperature $T_{\rm rh}$ is not affected by this interplay between $n_t$ and $r$; but we observe a more pronounced peak in the $T_{\rm rh}$ posterior distribution around $T_{\rm rh} \sim 100\,\textrm{MeV}$. Irrespective of whether our template for inflationary GWs is the correct one to describe the data or not, this means that the NG15 data in combination with the new noise models now show a stronger preference for a spectral break in the NANOGrav frequency band at a frequency $f \lesssim 14/T_{\rm obs}$. This observation is consistent with our results in Fig.~\ref{fig:bpl30$f$}, i.e., the posteriors for the broken power law, according to which the new pulsar noise models result in larger posterior support for a break frequency $f_b$ between the 8th and the 14th frequency bin than the old pulsar noise models. In the model for inflationary GWs, this break corresponds to the transition from the reheating phase to the era of radiation domination in the transfer function for primordial tensor modes. 

In the case of the phase-transition model, we observe shifts to lower values in the posterior distributions of all three parameters: the temperature scale of the phase transition, $T_*$; the strength of the phase transition, $\alpha$; and the radius of the colliding vacuum bubbles in units of the Hubble radius, $H_* R_*$. This trend indicates a preference for a weaker phase transition, which is consistent with our observation of an overall weaker GW signal in the CNM framework. We also note that the $95\,\%$ credible interval for $T_*/\textrm{MeV}$ in the HD analysis changes from $[4.0,110]$ to $[1.4,52]$, with the MAP value moving from $T_*^{\rm MAP} \simeq 18\,\textrm{MeV}$ to $T_*^{\rm MAP} \simeq 4.7\,\textrm{MeV}$. This new posterior range is further away from the temperature scale of the quantum chromodynamics (QCD) crossover in the Standard Model, $T_{\rm QCD} \sim 160\,\textrm{MeV}$, and closer to the onset of big-bang nucleosynthesis (BBN) around $T_{\rm BBN} \sim 0.1\,\textrm{MeV}$, which warrants further scrutiny of the phase-transition interpretation from the perspective of cosmological constraints~\citep{Bai:2021ibt} and the construction of microscopic models~\citep{Bringmann:2026xcx}. 

Finally, let us comment on our results for stable cosmic strings in Fig.~\ref{fig:strings}. In the language of \cite{ng15newphysics}, this scenario corresponds to the \textsc{stable-n} model, which models GW emission from closed loops based on the numerical simulations in \cite{Blanco-Pillado:2017oxo}. For DMX, the CURN analysis results in a MAP cosmic-string tension of $\log_{10}(G\mu)^{\rm MAP} \simeq -10.08$ (see the orange dashed posterior in Fig.~\ref{fig:strings}). This analysis can be improved in two ways: the CURN free spectrum can be replaced by the HD free spectrum, and DMX can be replaced by CNM. Interestingly enough, either of these corrections shifts the MAP cosmic-string tension by $\Delta \log_{10}G\mu \simeq 0.12$ to $\log_{10}(G\mu)^{\rm MAP} \simeq -10.20$ (see the solid orange and dashed blue posteriors in Fig.~\ref{fig:strings}). At the same time, the $\log_{10}G\mu$ distribution for DMX HD is slightly broader than for CNM CURN. Finally, applying also the respectively other correction yields an extra shift by $\Delta \log_{10}G\mu \simeq 0.07$, leading us to a final value of $\log_{10}(G\mu)^{\rm MAP} \simeq -10.27$ for CNM HD (see the blue solid posterior in Fig.~\ref{fig:strings}). We thus find a $35\%$ decrease in the best-fit value for the cosmic-string tension $G\mu$, which implies a $20\%$ reduction of the energy scale of the string-producing phase transition, $v \propto (G\mu)^{1/2}$. The same trend is reflected in the $95\%$ credible intervals that shift from $[-10.34,-9.78]$ (DMX CURN) over $[-10.46,-9.85]$ (DMX HD) and $[-10.43,-9.92]$ (CNM CURN) to $[-10.52,-10.02]$ (CNM HD). These results highlight that precision parameter inference for exotic spectral models requires accurate modeling of both the GWB and pulsar noise. For cosmic strings, we find that corrections to both model components can have a comparable impact on the inferred new-physics parameters. 

\section{Conclusions}
\label{sec:conclusions}

In this work we quantified the impact of the latest noise models for chromatic propagation effects from \citet{NG15_CNM} on a variety of past GW analyses using the NG15 dataset \citep{ng15data, ng15gwb}. In undertaking this, we accomplish two main goals: 1) to provide an updated assessment of current nHz GW inferences from NG15, and 2) to inform analyses of future NANOGrav datasets which may use similar noise modeling techniques.

In a broad sense, the CNMs provide dedicated channels for chromatic signals which may otherwise contaminate the analysis across the nHz band. They also reduce the overall prior volume of the model as compared with DMX (which otherwise accurately models DM variability, \citealt{Lentati+2017:wideband, Iraci+2024, Larsen2024}). This results in broadband improvements to GW sensitivity as confirmed by our sensitivity curve analyses, especially at high nHz frequencies. As such, the CNMs yield a new HD-correlated GWB spectrum (Fig.~\ref{fig:free_spec_comp}), which carries information on the underlying GW sources. 



Concretely, we are more confident in this updated spectral characterization of the GWB because the CNMs improve the statistical evidence for the HD correlations of the GWB. Comparing the Bayes Factor for HD correlations over a CURN under the ``flagship'' 14$f$ power-law spectral model from \citet{ng15gwb}, we find $\mathcal{B}^{\rm HD}_{\rm CURN} = 1571 \pm 14$ under the CNMs, roughly eight times larger than previously found in \citet{ng15gwb}. 
Likewise, reproducing analyses using the PTA optimal statistic \citep{ng15gwb, ng15_ppc} yields an improvement ${\rm S/N} = 5 \to 5.3$, corresponding to a reduction of the Bayesian $p$-value $p_B = 7.9 \times 10^{-4} \to 4.5 \times 10^{-4}$ when applying our CNMs. Our per-frequency analysis of the Bayes Factor (Fig.~\ref{fig:bf_vs_f}) strongly suggests that this improved evidence comes from improved noise mitigation 
in the 10 to 30 nHz range. With better-resolved cross-correlations under our CNMs, we also tighten upper limit constraints from \citet{ng15_altpol} on the amplitude of a breathing mode of gravity, which modifies the shape of the HD curve.


Improved GWB spectral characterization has implications on the underlying population of GW sources. Firstly, the consistency of the GWB with an astrophysical SMBHB population is improved with the measurement of a higher power-law spectral index closer to $\gamma = 13/3$. This result is consistent with the original ``DMGP'' analysis from \citet{ng15gwb} and other previously observed impacts of improved pulsar noise modeling \citep{pptadr3:noise, Goncharov+2025, ng12p5_CNM}. Reduction of the GWB spectral amplitude $A$ at fixed-$\gamma$ further implies fewer or less massive GW sources. Accounting for more detailed astrophysics using a phenomenological SMBHB population model from \citet{ng15smbbh}, this translates to only a slight reduction of the galaxy characteristic mass $m_{\psi,0}$ posterior. As such, we expect the isolated impact of these noise models will be to ease, but not entirely resolve, various tensions between the measured nHz GWB and galaxy population models (e.g., \citealt{sato-polito2024, Mingarelli2026}). Cosmological interpretations of the GWB from exotic physics in the early Universe \citep{ng15newphysics} are similarly impacted, e.g., with a weaker phase-transition signal or a reduced cosmic string tension.

Finer changes in spectral features at low GW frequencies ($f < 30$ nHz) include a slight preference for a turnover in the 2 to 4 nHz range of the GWB spectrum. While the turnover itself is not yet statistically significant, this implies e.g., less time spent in the stellar scattering phase of SMBHB hardening assuming an astrophysical GWB, or a more pronounced reheating phase if assuming an inflationary GWB. Reproducing the all-sky CW searches from \citet{ng15singlesource} shows increased evidence for CW power at $\sim4.2$ nHz, also identified by \citet{eptadr2_4:single_source}. Nonetheless, the power remains insignificant while assuming an HD-correlated spectrum, with $\mathcal{B}^{\rm HD+CW}_{\rm HD} = 1.69\pm0.06$. Meanwhile, a previously identified excursion from the power-law GWB at 16 nHz \citep{ng15-discreteness}, is no longer significant under our CNMs, specifically due to improvements in PSR J1713+0747's modeling \citep{lam+2018, NG15_CNM}. We also reproduce \emph{targeted} CW searches for SMBHB signals from two select candidate galaxies at 14 nHz (SDSS J0729+4008) and 21 nHz (SDSS J1536+0441), whose Bayes Factors were outliers among 114 AGN \citep{ng15-targeted}. We find the CNMs reduce the statistical significance of these CW signals down to the levels of the remaining population. That said, other candidates from \citet{ng15-targeted} not investigated here may emerge with stronger evidence under CNMs.

While the search for nHz GWs at high frequencies ($f > 30$ nHz) remains challenging, especially considering the lack of expected signals, CNMs provide substantial gains in sensitivity in this regime. We reduce the previous upper limit \citep{ng15-targeted} on the chirp mass of SMBHB candidate 3C 66B from $\mathcal{M}_c^{95\%} = 1.06(3) \times 10^9 M_{\odot}$ to $\ulcnmhdthreec$ $M_{\odot}$, assuming a circular binary emitting at $f_{\rm gw} \sim 60.4$ nHz.
In contrast, some of our results using the stochastic GWB models are more surprising: our per-frequency analysis suggests the presence of HD-correlated power even above $f > 30$ nHz, and the high-frequency flattening of the CURN spectrum becomes more pronounced under the CNMs, contrary to previous expectations from \citet{ng12p5_CNM}. While interesting, this high-frequency power is not statistically significant. As such, it may be consistent with the presence of unresolved CW power at high frequencies, but just as likely may be the result of unaccounted systematics such as remaining imperfections in the pulsar noise models or spectral leakage artifacts \citep{Crisostomi+2025}. 

Looking to future datasets, it is imperative to continue pursuing detailed modeling of chromatic effects in order to optimize detector sensitivity to nHz GWs and ensure their accurate characterization. While we have employed a customized noise modeling approach in \citet{NG15_CNM}, a comprehensive accounting for any previously known or expected effects should take first priority over statistical and computational optimization in future work \citep{vH2025_model_averaging, DiMarco+2025}. This pursuit will complement other recent advances in GW analysis methods, such as improved timing methods \citep{Susobhanan+2025_vela, Curylo+2026}, improved achromatic noise and GW models \citep{Crisostomi+2025, Deng+2026}, improved sampling of the full-PTA parameters \citep{Vallisneri+2025, Valtolina2025, Laal+2025_solving_PTA, Gundersen+2025}, and pulsar-population-informed priors \citep{vanHaasteren2024, GoncharovSardana2025}. The benefits of these methodologies will only compound with the promise of improved timespans, bandwidths, and TOA precisions from future PTA datasets.


\section*{Acknowledgements}
The NANOGrav collaboration receives support from National Science Foundation (NSF) Physics Frontiers Center award \#2020265, the Gordon and Betty Moore Foundation, an NSERC Discovery Grant, and CIFAR.
The Arecibo Observatory is a facility of the NSF operated under cooperative agreement (AST-1744119) by the University of Central Florida (UCF) in alliance with Universidad Ana G. M\'endez (UAGM) and Yang Enterprises (YEI), Inc.
The Green Bank Observatory is a facility of the NSF operated under cooperative agreement by Associated Universities, Inc.
The National Radio Astronomy Observatory is a facility of the NSF operated under cooperative agreement by Associated Universities, Inc.
J.G.B.\ was supported in part through NASA and Oregon Space Grant Consortium, cooperative agreement 80NSSC20M0035.
J.G.B., J.S.H., and K.W.\ acknowledge support from NSF CAREER award \#2339728.
Y.C. was supported in part by NASA CT Space Grant PTE Federal Award Number 80NSSC20M0129, as well as the the Yale College Dean’s Office and the STARS Program.
C.M.F.M.\ was supported in part by the National Science Foundation under Grants No.\ NSF PHY-1748958,  AST-2106552, and NASA LPS 80NSSC24K0440. C.M.F.M. also thanks the Center for Computational Astrophysics (CCA) of the Flatiron Institute for support. The Flatiron Institute is supported by the Simons Foundation.
J.S.\ is supported by an NSF Astronomy and Astrophysics Postdoctoral Fellowship under award AST-2202388, and acknowledges previous support by the NSF under award 1847938.
D.~H.~A.\ acknowledges support from Studienstiftung des Deutschen Volkes.
K.~Sc.\ is an affiliate member of the Kavli Institute for the Physics and
Mathematics of the Universe (Kavli IPMU) at the University of Tokyo and
supported by the World Premier International Research Center Initiative
(WPI), MEXT, Japan (Kavli IPMU).
L.B.\ acknowledges support from the National Science Foundation under award AST-2307171 and from the National Aeronautics and Space Administration under award 80NSSC22K0808.
P.R.B.\ is supported by the Science and Technology Facilities Council, grant number ST/W000946/1.
S.B.\ gratefully acknowledges the support of a Sloan Fellowship, and the support of NSF under award \#1815664.
The work of R.B., R.C., X.S., J.T., and D.W.\ is partly supported by the George and Hannah Bolinger Memorial Fund in the College of Science at Oregon State University.
M.C.\ acknowledges support by the European Union (ERC, MMMonsters, 101117624).
Support for this work was provided by the NSF through the Grote Reber Fellowship Program administered by Associated Universities, Inc./National Radio Astronomy Observatory.
H.T.C.\ acknowledges funding from the U.S. Naval Research Laboratory.
Pulsar research at UBC is supported by an NSERC Discovery Grant and by CIFAR.
K.C.\ is supported by a UBC Four Year Fellowship (6456).
M.E.D.\ acknowledges support from the Naval Research Laboratory by NASA under contract S-15633Y.
T.D.\ and M.T.L.\ received support by an NSF Astronomy and Astrophysics Grant (AAG) award number 2009468 during this work.
E.C.F.\ is supported by NASA under award number 80GSFC24M0006.
K.A.G.\ and S.R.T.\ acknowledge support from an NSF CAREER award \#2146016.
D.C.G.\ is supported by NSF Astronomy and Astrophysics Grant (AAG) award \#2406919.
A.D.J.\ acknowledges support from the Caltech and Jet Propulsion Laboratory President's and Director's Research and Development Fund.
A.D.J.\ acknowledges support from the Sloan Foundation.
N.La.\ was supported by the Vanderbilt Initiative in Data Intensive Astrophysics (VIDA) Fellowship.
Part of this research was carried out at the Jet Propulsion Laboratory, California Institute of Technology, under a contract with the National Aeronautics and Space Administration (80NM0018D0004).
D.R.L.\ and M.A.M.\ are supported by NSF \#1458952.
M.A.M.\ is supported by NSF \#2009425.
C.M.F.M.\ was supported in part by the National Science Foundation under Grants No.\ NSF PHY-1748958 and NASA LPS 80NSSC24K0440. C.M.F.M.\ also thanks the Center for Computational Astrophysics (CCA) of the Flatiron Institute for support. The Flatiron Institute is supported by the Simons Foundation.
A.Mi.\ acknowledges support from a Royal Society University Research Fellowship (URF-R1-251896)
The Dunlap Institute is funded by an endowment established by the David Dunlap family and the University of Toronto.
K.D.O.\ was supported in part by NSF Grant No.\ 2207267.
T.T.P.\ acknowledges support from the Extragalactic Astrophysics Research Group at E\"{o}tv\"{o}s Lor\'{a}nd University, funded by the E\"{o}tv\"{o}s Lor\'{a}nd Research Network (ELKH), which was used during the development of this research.
P.P.\ and S.R.T.\ acknowledge support from NSF AST-2007993.
H.A.R.\ is supported by NSF Partnerships for Research and Education in Physics (PREP) award No.\ 2216793.
S.M.R.\ and I.H.S.\ are CIFAR Fellows.
Portions of this work performed at NRL were supported by ONR 6.1 basic research funding.
J.D.R.\ also acknowledges support from start-up funds from Texas Tech University.
S.C.S.\ and S.J.V.\ are supported by NSF award PHY-2011772.
J.S.\ is supported by an NSF Astronomy and Astrophysics Postdoctoral Fellowship under award AST-2202388, and acknowledges previous support by the NSF under award 1847938.
J.P.W.V.\ acknowledges support from NSF AccelNet award No.~2114721.
O.Y.\ is supported by the National Science Foundation Graduate Research Fellowship under Grant No.\ DGE-2139292.
A.Ash.\ is supported by the Gordon and Betty Moore Foundation and acknowledges previous support from the NSF PFC \#2020265.

\section*{Author Contributions}
All authors contributed to the activities of the NANOGrav collaboration leading to the work presented here and reviewed the manuscript, text, and figures prior to the paper’s submission. Additional specific contributions to this paper are as follows. J.G.B. and B.L. co-led the project, carried out the analyses, interpreted the results, and led the writing of the manuscript. J.S.H., C.M.F.M., and J.S. conceived of and advised on the project, and contributed to the interpretation of the analyses. Y.C., A.Ash., and D.J.O. contributed to carrying out analyses and interpreting results. C.Matt, C.J.H., W.G.L., K.G., and L.B. wrote and ran the analyses featured the astrophysical interpretations section. A.Am., D.H.A., R.S., and K.Sc. wrote and ran the analyses featured in the new physics interpretations section. J.G.B., B.L., D.J.O., K.W., Y.C., M.T.M., J.S.H., C.M.F.M., and J.S. meticulously developed, selected, and analyzed the custom noise models making this work possible. R.v.H. contributed the ranked-reduced framework for rapid computation of $p$-values.

The $15$-year data set was produced through a combination of observations, arrival time calculations, data checks and refinements, and timing-model development and analysis which were performed by the following authors: G.A., A.Anu., A.M.A., Z.A., P.T.B., P.R.B., H.T.C., K.C., M.E.D., P.B.D., T.D., E.C.F., W.F., E.F., G.E.F., N.G., P.A.G., J.G., D.C.G., J.S.H., R.J.J., M.L.J., D.L.K., M.K., M.T.L., D.R.L., J.L., R.S.L., A.M., M.A.M., N.M., B.W.M., C.N., D.J.N., T.T.P., B.B.P.P., N.S.P., H.A.R., S.M.R., P.S.R., A.S., C.S., B.J.S., I.H.S., K.S., A.S., J.K.S., H.M.W.

\facilities{Arecibo, GBT, VLA}

\software{
    \texttt{astropy} \citep{astropy},
    \texttt{numpy} \citep{numpy},
    \texttt{scipy} \citep{scipy},
    \texttt{pandas} \citep{pandas},
    \texttt{PINT} \citep{PINT}, 
    \texttt{enterprise} \citep{enterprise},
    \texttt{enterprise\_extensions} \citep{enterprise_ext},
    \texttt{PTMCMCSampler} \citep{ptmcmcsampler},
    \texttt{la\_forge} \citep{laforge2020},
    \texttt{hasasia} \citep{hazboun:2019has},
    \texttt{defiant} \citep{Gersbach+2025},
    \texttt{ceffyl} \citep{lamb2023rapid},
    \texttt{holodeck} \citep{ng15smbbh},
    \texttt{PTArcade} \citep{PTArcade},
    \texttt{QuickCW} \citep{becsy2022quickCW},
    \texttt{matplotlib} \citep{matplotlib},
    \texttt{kalepy} \citep{kalepy},
    \texttt{healpy} \citep{healpy},
    \texttt{corner} \citep{corner}
}

\appendix

\section{Table of priors}
\label{appendix:priors}

\begin{table*}[ht]
\begin{center}
\scriptsize
\label{tab:priors}
\begin{tabular}{llll}
\hline\hline
parameter & description & prior & comments \\
\hline

\multicolumn{4}{c}{intrinsic red noise} \\[1pt]
$A_{\rm red}$ & red-noise power-law amplitude& log-Uniform $[-20, -11]$ & one parameter per pulsar  \\
$\gamma_{\rm red}$ & red-noise power-law spectral index & Uniform $[0, 7]$ & one parameter per pulsar \\
\hline

\multicolumn{4}{c}{all common processes, free spectrum} \\[1pt]
$\rho_{i}$ [s$^{2}$] & power-spectrum coefficients at $f=i/T$ & log-Uniform in $\rho_{i}$ $[-18,-8]$ & one parameter per frequency\\
\hline

\multicolumn{4}{c}{all common processes, power-law spectrum} \\[1pt]
$A$ & common process strain amplitude & log-Uniform $[-18, -14]$ ($\gamma=13/3$) & one parameter for PTA \\
& & log-Uniform $[-18, -11]$ ($\gamma$ varied) & one parameter for PTA \\
$\gamma$ & common process power-law spectral index & delta function ($\gamma=13/3$)& fixed \\
& & Uniform $[0,7]$ & one parameter for PTA \\
\hline

\multicolumn{4}{c}{all common processes, broken--power-law spectrum} \\[1pt]
$A$ & broken--power-law amplitude & log-Uniform $[-18, -11]$ & one parameter for PTA \\
$\gamma$ & broken--power-law low-freq.\ spectral index & Uniform $[0,7]$ & one parameter per PTA \\
$\delta$ & broken--power-law high-freq.\ spectral index & delta function ($\delta=0$) & fixed \\
$f_{\rm bend}$ [Hz] & broken--power-law bend frequency & log-Uniform  [$-8.7$,$-7$] & one parameter for PTA \\
$\ell$ & broken--power-law high-freq.\ transition sharpness & delta function ($\ell=0.1$) & fixed \\
\hline

\multicolumn{4}{c}{CW Searches} \\[1pt]
$A$ & common process strain amplitude & log-Uniform $[-18, -11]$ & either HD or CURN as noted \\
$\gamma$ & common process power-law spectral index & Uniform $[0, 7]$ & either HD or CURN as noted \\
$\log_{10} h$ & CW strain amplitude & log-Uniform $[-18, -11]$ & implicitly fixed for targeted searches \\
$\log_{10} \mathcal{M}_c$ & chirp mass & Uniform $[7, 11]$ & \\
$\log_{10} d_L$ & luminosity distance & -- & fixed for targeted searches \\
$\log_{10} f_{\rm GW}$ & GW frequency & Uniform $[-9, -6.52]$ & fixed for targeted searches \\
$\phi_{\rm GW}$ & ecliptic longitude & Uniform $[0, 2\pi]$ & fixed for targeted searches \\
$\cos\theta_{\rm GW}$ & ecliptic colatitude & Uniform $[-1, 1]$ & fixed for targeted searches \\
$\psi$ & polarization angle & Uniform $[0, \pi]$ & \\
$\Phi_0$ & initial GW phase & Uniform $[0, 2\pi]$ & \\
$\cos\iota$ & inclination angle & Uniform $[-1, 1]$ & \\
\hline

\multicolumn{4}{c}{Astrophysical Interpretation} \\[1pt]
$\psi_0$ & GSMF normalization & Uniform $[-3.5, -1.5]$ & \\
$m_{\psi,0}$ & GSMF characteristic mass & Uniform $[10.5, 12.5]$ & \\
$\mu$ & $M_{\rm BH}$--$M_{\rm bulge}$ normalization & Uniform $[7.6, 9.0]$ & \\
$\epsilon_\mu$ & $M_{\rm BH}$--$M_{\rm bulge}$ scatter & Uniform $[0.0, 0.9]$ dex & \\
$\tau_f$ & total binary lifetime & Uniform $[0.1, 11.0]$ Gyr & \\
$\nu_{\rm inner}$ & small separation hardening power-law index & Uniform $[-1.5, 0.0]$ & \\
\hline

\multicolumn{4}{c}{New Physics} \\[1pt]
$\log_{10}T_{\mathrm{rh}}/\mathrm{GeV}$ & reheating temperature & Uniform $[-3, 3]$ & one parameter per PTA \\
$\log_{10}r$ & tensor-to-scalar ratio & Uniform $[-40, 0]$ & one parameter per PTA \\
$n_t$ & tensor spectral index & Uniform $[0, 6]$ & one parameter per PTA \\
\noalign{\vskip 5pt}
$\log_{10}T_{\star}/\mathrm{GeV}$ & percolation temperature & Uniform $[-4, 4]$ & one parameter per PTA \\
$\log_{10}{\alpha}_{\star}$ & strength of the phase transition & Uniform $[-2, 1]$ & one parameter per PTA \\
$\log_{10}H_{\star}R_{\star}$ & bubble radius in units of the Hubble radius & Uniform $[-3, 0]$ & one parameter per PTA \\
\noalign{\vskip 5pt}
$\log_{10}G\mu$ & cosmic-string tension & Uniform $[-14, -6]$ & one parameter per PTA \\
\hline

\end{tabular}
\caption{Prior distributions used in various analyses in this work. Note that when we allow chromatic parameters to vary during the analysis, several additional parameters are introduced per pulsar, for which the priors can be found in the appendix of \citet{NG15_CNM}.}
\end{center}
\end{table*}


\bibliographystyle{aasjournal}
\bibliography{hazgrav}

\end{document}